\documentclass[journal, amsmath,amssymb]{revtex4-1}
\usepackage{graphicx}
\usepackage{bm}
\usepackage{xcolor}
\begin{document}
\title{Motion of a parametrically driven damped coplanar double pendulum}
\author{Rebeka Sarkar}
\affiliation{Department of Physics, Indian Institute of Technology Kharagpur, Kharagpur-721302, West Bengal, India}
\author{Krishna Kumar}
\thanks{E-mail: kumar.phy.iitkgp@gmail.com}%
\affiliation{Department of Physics, Indian Institute of Technology Kharagpur, Kharagpur-721302, West Bengal, India}
\author{Sugata Pratik Khastgir}
\affiliation{Department of Physics, Indian Institute of Technology Kharagpur, Kharagpur-721302, West Bengal, India} 
\date{\today} 
\begin{abstract}
We present the results of linear stability of a damped coplanar double pendulum and its non-linear motion, when the point of suspension is vibrated sinusoidally in the vertical direction with amplitude $a$ and frequency $\omega $. A double pendulum has two pairs of Floquet multipliers, which have been calculated for various driving parameters. We have considered the stability of a  double pendulum when it is in any of its possible stationary states: (i) both pendulums are either vertically downward or upward and (ii) one pendulum is downward and other is upward. The damping is considered to be velocity-dependent, and the driving frequency is taken in a wide range. A double pendulum excited from its stable state shows both periodic and chaotic motion. The periodic motion about its pivot may be either oscillatory or rotational. The periodic swings of a driven double pendulum may be either harmonic or subharmonic for lower values of $a$. The limit cycles corresponding to the normal mode oscillations of a double pendulum of two equal masses are squeezed into a line in its configuration space. For unequal masses, the pendulum shows multi-period swings for smaller values of $a$ and damping, while chaotic swings or rotational motion at relatively higher values of $a$. The parametric driving may lead to stabilization of a partially or fully inverted double pendulum. 
\end{abstract}
\pacs{Valid PACS appear here}
\maketitle
	

\section{\label{sec:Introduction}Introduction}

Parametric excitation of surface waves in a fluid is known since the famous experiment by Faraday~\cite{Faraday_1831,Benjamin-Ursel_1954,Douady_1990,Miles-Henderson_1990}. These standing waves, also known as Faraday waves~\cite{Ciliberto-Gollub_1984,Tuffilaro_etal_1989,Fauve_etal_1992,Mueller_1993,Edwards-Fauve_1993,Kumar-Tuckerman_1994,Edwards-Fauve_1994,Kumar-Bajaj_1995,Gollub_1995,Maity_etal_2020,Lau_2020}, oscillate with a frequency equal to half the driving frequency. The surface waves in a thin layer of viscous liquid can also be synchronous (harmonic) with the external driving~\cite{Kumar_1996,Mueller_etal_1997} at the instability onset. Interfacial patterns in a Faraday experiment with relatively low-viscosity fluids~\cite{Douady_1990,Ciliberto-Gollub_1984,Tuffilaro_etal_1989} are known to differ significantly from those observed in moderate or high-viscosity fluids~\cite{Fauve_etal_1992,Mueller_1993,Edwards-Fauve_1993,Kumar-Tuckerman_1994,Edwards-Fauve_1994,Kumar-Bajaj_1995,Gollub_1995}.
The phenomenon of parametric resonance has been attracting attention in diverse areas of research. They include condensed matter physics~\cite{Engelss-Atherton_2007,Nguyen_etal_2019}, optics~\cite{Leonardo_etal_2007,Svidzinsky_etal_2013,Zhang_etal_2017,Nessler_etal_2020}, cosmology~\cite{Kofman_etal_1996}, quantum field theory~\cite{Berges_Serreau_2003}, non-linear science~\cite{Ripoll_Gracia_1999,Rowland_2004}, biophysics~\cite{Maksymov_Pototsky_2020}, modulated fluid flows~\cite{Smorodin_Luecke_2010,Kaur_etal_2016,Hazra_etal_2020} etc. 

A parametrically driven planar pendulum~\cite{Stephenson_1908,Kapitza_1951,Mechanics_Landau-Lifshitz_1960,Mclaughlin_1981,Leven_etal_1985,Acheson_1995,Starrett-Tagg_1995,CliffordBishop_1998,Sang-Kim_1998,Bartucelli_2001,Butikov_2001} may oscillate either sub-harmonically or harmonically or rotate about its point of support.  However, the critical value of driving amplitude to excite subharmonic swing in the presence of damping is always minimum. The motion of a planar pendulum is restrictive due to its single degree of freedom. However, a double pendulum, with a couple of degrees of freedom, has enlarged phase space. A parametrically driven double pendulum has the potential to capture several interesting phenomena observed in fluids and in other coupled systems under parametric driving. The double pendulum with two equal masses in the absence of damping is a well-studied system.  It is one of the classic problems of mechanics that has attracted the attention of scientists and engineers for more than a century. The swinging motion of a double pendulum~\cite{Mechanics_Landau-Lifshitz_1960,Stephenson_1908} shows a rich dynamics~\cite{Shinbrot_etal_1992,Levien-Tan_1993,Stachowiak-Okada_2000,Cross_2011,Acheson_1993,Acheson-Mullin_1993,Mullin_etal_2003}. As it has two normal modes, it displays exciting and complex oscillatory as well as rotational motion. If the point of suspension is vibrated sinusoidally in the vertical direction, then a damped double pendulum is expected to display either periodic or aperiodic motion, if the driving amplitude is raised above a threshold. In addition, the linearized equations of motion of a double pendulum lead to a system of coupled Mathieu equations~\cite{Mathieu_1868}, which is used as a model to study several interesting phenomena~\cite{Nguyen_etal_2019,Leonardo_etal_2007,Svidzinsky_etal_2013,Zhang_etal_2017,Nessler_etal_2020,Kofman_etal_1996,Berges_Serreau_2003}. The nonlinear motion is more complex. 

In this paper, we study the effects of (1) damping and (2) unequal masses on the motions of a double pendulum. It is observed that richer dynamics emerge when one incorporates the two above-mentioned features in the system. The understanding of linear stability and non-linear motion of a parametrically driven double pendulum, in the presence of damping, is useful for the robotics and sports industries. The linear stability of a damped double pendulum under parametric driving is not known if the two masses are unequal. We present here the results of numerical studies on the motion of a parametrically driven damped coplanar double pendulum. The Floquet multipliers for the pendulum are also computed in the absence of damping. The linear stability using the Floquet method shows a qualitatively new phenomenon, where two marginal stability curves for unequal masses merge together in the space of driving parameters to form a double-well-shaped new instability zone. The non-linear swings and rotations of the pendulum are investigated by integrating the equations of motion numerically. The limit cycles corresponding to normal mode oscillations are squeezed to a line in the configuration space. The Lyapunov exponents and the phase portraits are also computed numerically. The stability of the system with one or both the pendulums inverted is also investigated. The nonlinear motion of a partially inverted double pendulum is also studied, which shows interesting features. 	
\maketitle
	
\section{Physical system and the equations of motion}
\begin{figure}[h!]
\centering
\includegraphics[height=!, width=8.5cm] {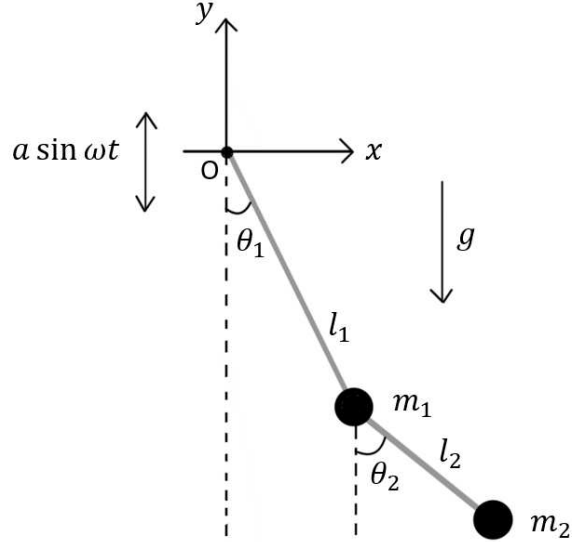}
\caption{Schematic diagram of a parametrically driven coplanar double pendulum. The pivot is vibrated sinusoidally in the vertical direction.}
\label{fig:schematic}
\end{figure}

We consider a coplanar damped double pendulum, whose pivot is vibrated sinusoidally in the vertical direction. A mass $m_1$ connected to a light rod of length $l_1$ is suspended from a fixed pivot. Another mass $m_2$ connected to a light rod of length $l_2$ is attached to the first one, as shown in Fig.~\ref{fig:schematic} schematically below. When at rest, both the masses are below the pivot along a vertical line passing through the pivot. The upper pendulum may oscillate or rotate about the pivot, while the lower one may do the same about its point of suspension. The dissipative forces $Q_1$ and $Q_2$ acting on the upper and lower pendulums are considered to be velocity dependent, and derivable from the Rayleigh dissipation function~\cite{Goldstein}, $\mathcal{F} = \frac{\Gamma}{2} (\mathrm{v}_1^2 + \mathrm{v}_2^2)$, where $\Gamma$ is the damping coefficient and $\mathbf{v}_i$ is the instantaneous velocity of the bob of the $i$th pendulum. A two-dimensional coordinate system is chosen with its origin at the mean position of the vibrating pivot. The $x$ and $y$- axes are along the horizontal and vertical directions through the origin. The pendulum of length $l_i$ and mass $m_i$ ($i = 1, 2$) makes an angle $\theta_i$ with the vertical direction, as shown in the figure.
In terms of generalized coordinates ($\theta_1$ and $\theta_2$) and parameters of the double pendulum, the position vectors of the two masses are
\begin{eqnarray}
	\mathbf{r}_1 (t) &=& l_1 \sin{\theta_1} \hat{\mathbf{x}} + (-l_1 \cos{\theta_1} + a \sin{\omega t}) \hat{\mathbf{y}}\\
	\mathbf{r}_2 (t) &=& \mathbf{r}_1 (t) + l_2 (\sin{\theta_2}\hat{\mathbf{x}}  - \cos{\theta_2}\hat{\mathbf{y}}),
\end{eqnarray}
where $\hat{\mathbf{x}}$ and $\hat{\mathbf{y}}$ are unit vectors along the positive directions of the $x$ and $y$ axes, respectively.
The Lagrangian function $\mathcal{L}(\theta_1, \theta_2, \dot{\theta}_1, \dot{\theta}_2, t)$ of the system reads as~\cite{Mechanics_Landau-Lifshitz_1960}:
\begin{eqnarray}	
	\mathcal{L}&=&\frac{1}{2} (m_1 + m_2) l_1^2 \dot{\theta}_1^2 + \frac{1}{2} m_2 l_2^2 \dot{\theta}_2^2  
	+ (m_1 + m_2) g l_1 \cos{\theta_1}
	+  m_2 g l_2\cos{\theta}_2 
	+  m_2 l_1 l_2 \dot{\theta}_1\dot{\theta}_2 \cos{(\theta_2-\theta_1)} \nonumber \\
	&+& (m_1 + m_2) a \omega l_1 \dot{\theta}_1 \sin{\theta_1}\cos{\omega t}
	+ m_2 a \omega l_2 \dot{\theta}_2 \sin{\theta_2} \cos{\omega t}  + \dot{f}_1(t), \label{Lag1}
\end{eqnarray}	
where $\dot{f}_1$ is total time derivative of an irrelevant function $f_1(t, a, \omega, m_1, m_2, g)$ in the Lagrangian. The Rayleigh dissipation function $\mathcal{F}(\theta_1, \theta_2, \dot{\theta}_1, \dot{\theta}_2, t)$ for parametrically driven coplanar pendulum may be written as
\begin{eqnarray}\label{eq:diss1}
	\mathcal{F}&=& \frac{\Gamma}{2} \left[ 2 l_1^2 \dot{\theta}_1^2 + l_2^2\dot{\theta}_2^2 + 2 l_1 l_2 \dot{\theta}_1 \dot{\theta}_2 \cos{(\theta_1 - \theta_2)} \right. 
	+ \left. 2 a \omega \left( 2 l_1  \dot{\theta}_1 \sin{\theta_1} \cos{\omega t} +  l_2 \dot{\theta}_2 \sin{\theta}_2 \cos{\omega t} \right) \right] \nonumber\\ 
	&+& f_2(t),
\end{eqnarray}
where the function $f_2(t) = \Gamma a^2 \omega^2 \cos^2{\omega t}$ is independent of generalized coordinates and velocities. The dissipative forces $Q_1$ and $Q_2$, in terms of the generalized coordinates and velocities, acting on masses $m_1$ and $m_2$, respectively, are

\begin{eqnarray}
	Q_1 = -\frac{\partial \mathcal{F}} {\partial  \dot{\theta}_1} 
	&=& -\Gamma{\Big [} 2l_1^2\dot{\theta}_1 + l_1 l_2\dot{\theta}_2 \cos{(\theta_1-\theta_2)} 
	+ 2a \omega l_1 \sin{\theta_1} \cos{\omega t}{\Big ]}, \label{damp1} 
\end{eqnarray}	
\begin{eqnarray}
	Q_2 = -\frac{\partial \mathcal{F}} {\partial  \dot{\theta}_2} &=&-\Gamma{\Big [} l_2^2\dot{\theta}_2 + l_1 l_2\dot{\theta}_1 \cos{(\theta_1-\theta_2)} 
	+ a \omega l_2\sin{\theta_2} \cos{\omega t}{\Big ]}. \label{damp2}
\end{eqnarray}
The equations of motion for a coplanar double pendulum may be derived using the Euler-Lagrange equations, as follows: 
\begin{equation}
	\frac{d}{dt}\left(\frac{\partial \mathcal{L}} {\partial \dot{\theta}_i}\right)-\frac{\partial \mathcal{L}}{\partial \theta_i} = Q_i;~~ i= 1, 2.\label{EL}
\end{equation}
Inserting the expressions for Lagrangian (Eq.~\ref{Lag1}) and dissipative forces [Eqs.~(\ref{damp1}-\ref{damp2})] in Eq.~(\ref{EL}), we get the following coupled equations for the driven double pendulum.

	\begin{eqnarray}
	&&(m_1+m_2)l_1^2\Ddot{\theta}_1 + (m_1 + m_2) g l_1 \left(1-\frac{a\omega^2}{g}\sin{\omega t}\right)  \sin{\theta_1} +  m_2 l_1 l_2 \dot{\theta}_2^2 \sin{(\theta_1-\theta_2)}  \nonumber \\ 
	&+& \Gamma \left[ 2l_1^2 \dot{\theta}_1 +  l_1 l_2 \dot{\theta}_2 \cos{(\theta_1-\theta_2)} + 2a \omega l_1\sin{\theta_1} \cos{\omega t} \right] = - m_2 l_1 l_2 \Ddot{\theta}_2  \cos{(\theta_1-\theta_2)}, \label{EL1}
\end{eqnarray}
\begin{eqnarray}
	&&m_2l_2^2\Ddot{\theta}_2 + m_2gl_2 \left( 1 - \frac{a \omega^2}{g} \sin{\omega t} \right) \sin{\theta_2} -  m_2 l_1 l_2 \dot{\theta}_1^2 \sin{(\theta_1-\theta_2)} \nonumber \\
	&+& \Gamma\left[l_2^2 \dot{\theta}_2 + l_1 l_2 \dot{\theta}_1 \cos{(\theta_1-\theta_2)} + a \omega l_2 \sin{\theta_2} \cos{\omega t}\right] = - m_2l_1 l_2 \Ddot{\theta}_1 \cos{(\theta_1-\theta_2)}. \label{EL2}
\end{eqnarray}
The equation for $\Ddot{\theta}_1$ involves  $\Ddot{\theta}_2$ and vice versa. One may rewrite these equations where the equation for $\Ddot{\theta}_1$ ($\Ddot{\theta}_2$) does not involve directly $\Ddot{\theta}_2$ ($\Ddot{\theta}_1$). We do that after making the equations of motion [Eqs.~(\ref{EL1}) and (\ref{EL2})] dimensionless. We define dimensionless time  $\tau = \omega_0 t$, where $\omega_0 = \sqrt{g/l_1}$. We also introduce the following dimensionless parameters: $\mu = m_2/m_1$ for the mass ratio, $\lambda = l_1/l_2$ for the length ratio,  $2\beta = \Gamma/(m_1 \omega_0)$ for the damping coefficient. The dimensionless driving amplitude $A$ and the frequency $\Omega$  are defined as  $A= (a \omega^2)/g$ and $\Omega = \omega/\omega_0$, respectively. The equations of motion may now be written as 
	\begin{eqnarray}
	&&\left[ 1 + \mu \sin^2{(\theta_1-\theta_2)}\right] \Ddot{\theta}_1 +  (1-A\sin{\Omega \tau}) \left[ (1+\mu)\sin{\theta_1}-\mu \cos{(\theta_1-\theta_2)}\sin{\theta_2}\right] \nonumber \\
	&&+ \frac{\mu}{\lambda} \left[ \dot{\theta}_2^2 + \lambda \dot{\theta}_1^2 \cos{(\theta_1 -\theta_2)} \right] \sin{(\theta_1 -\theta_2)} + 2\beta \left[ 1 +\sin^2{(\theta_1-\theta_2)} \right] \dot{\theta}_1 \nonumber \\
	&&+\frac{2\beta}{\Omega} A \left[ 2\sin{\theta_1}-\cos{(\theta_1-\theta_2)}\sin{\theta_2}\right] \cos{\Omega \tau} = 0, \label{eq:dp1}
\end{eqnarray}
\begin{eqnarray}
	&&\left[ 1 + \mu \sin^2{(\theta_1-\theta_2)}\right] \Ddot{\theta}_2 + \lambda (1+\mu) (1-A\sin{\Omega \tau}) \left[\sin{\theta_2} -\cos{(\theta_1 -\theta_2)} \sin{\theta_1}\right]\nonumber\\
	&&- \left[ \lambda (1+\mu)\dot{\theta}_1^2 +\mu \dot{\theta}_2^2 \cos{(\theta_1-\theta_2)} \right] \sin{(\theta_1-\theta_2)}\nonumber \\
	&&+ \frac{2\beta}{\mu} \left[ \left\{ 1 + \mu \sin^2{(\theta_1-\theta_2)} \right\} \dot{\theta}_2+\lambda (1-\mu)\dot{\theta}_1 \cos{(\theta_1-\theta_2)} \right]\nonumber \\
	&&+ \frac{2\beta}{\Omega}\frac{\lambda}{\mu} A \left[ (1+\mu)\sin{\theta_2}-2\mu \cos{(\theta_1-\theta_2)} \sin{\theta_1}\right] \cos{\Omega \tau} = 0.\label{eq:dp2}
\end{eqnarray}
The set of above two second-order differential equations [Eqs.~(\ref{eq:dp1} and \ref{eq:dp2})] describes the dynamics of a parametrically driven coplanar double pendulum in the presence of velocity-dependent damping. These two equations are invariant under the transformation $\theta_1 \rightarrow -\theta_1$ and $\theta_2 \rightarrow -\theta_2$ (inversion symmetry) simultaneously for all values of $\lambda$, $\mu$ and $\beta$. Notice that the last terms in both the equations depend on $(\beta A/\Omega) \cos{\Omega \tau}$. They arise if the pivot is vibrated in the vertical direction. They affect both linear and non-linear motion of the parametrically driven double pendulum. If the two masses of the pendulum are equal ($\mu =1$), then Eqs.~(\ref{eq:dp1}) and (\ref{eq:dp2}) get simplified. They can be written as 
	\begin{eqnarray}
	&&\left[ 1 + \sin^2{(\theta_1-\theta_2)}\right] \left( \Ddot{\theta}_1 + 2 \beta \dot{\theta}_1 \right) +  [1-R\sin{(\Omega \tau - \Delta)}] \left[ 2\sin{\theta_1}-\cos{(\theta_1-\theta_2)}\sin{\theta_2}\right] \nonumber \\
	&&+  \left[ \frac{1}{\lambda} \dot{\theta}_2^2 + \dot{\theta}_1^2 \cos{(\theta_1 -\theta_2)} \right]\sin{(\theta_1-\theta_2)} = 0, \label{eq:dp1_mu1}
\end{eqnarray}
\begin{eqnarray}
	&&\left[ 1 + \sin^2{(\theta_1-\theta_2)}\right] \left( \Ddot{\theta}_2 + 2 \beta \dot{\theta}_2 \right) + 2 \lambda \left[ 1-R\sin{(\Omega \tau - \Delta)} \right] \left[\sin{\theta_2} -\cos{(\theta_1 -\theta_2)} \sin{\theta_1}\right]\nonumber\\
	&&- \left[ 2\lambda \dot{\theta}_1^2 + \dot{\theta}_2^2 \cos{(\theta_1-\theta_2)} \right] \sin{(\theta_1-\theta_2)}=0,
	\label{eq:dp2_mu1}
\end{eqnarray}
where $R = A \sec{\Delta} = A \sqrt{1 + \left( 2\beta/\Omega \right)^2}$ is the redefined driving amplitude and $\Delta$ $=$ $\tan^{-1}{(2\beta/\Omega)}$ is a phase factor.
	
\section{Linear stability analysis}
We now perform the Floquet stability analysis of a parametrically driven damped double pendulum when the point of support is vibrated vertically. The system has infinitely many fixed (equilibrium) points. They are denoted by ($\theta_1^*, \theta_2^*$) where $\theta_1^* = 0, \pm \pi, \pm 2\pi, \pm 3\pi, ...$ and $\theta_2^* = 0, \pm \pi, \pm 2\pi, \pm 3\pi, ...$.  However, each of these these fixed points is physically equivalent to one of the following four stationary states:\\ 
(i) Normal double pendulum: Both the pendulums are vertically downward ($\theta_1^* = \theta_2^* = 0$) initially.  This fixed point is stable if the amplitude of the external driving is below a threshold. \\ 
(ii) Partially inverted double pendulum: One of the two pendulums is inverted initially and the other is vertically downward.  The fixed point ($\theta_1^* = 0, \theta_2^* = \pi$) corresponds to a case when the pivoted pendulum is vertically downward and the other one is inverted. Another fixed point ($\theta_1^* = \pi, \theta_2^* = 0$) describes the stationary state when the pivoted pendulum is inverted and the second one is vertically downward. Both of these fixed points are unstable in the absence of external driving.\\
(iii) Fully inverted double pendulum: The fixed point ($\theta_1^* = \pi, \theta_2^* = \pi$) describes a stationary state when both the pendulums are initially inverted. This is also an unstable state in the absence of driving.\\
We consider the role of parametric driving on each of these four fixed points. The linearisation
of Eqs~(\ref{eq:dp1}) and~(\ref{eq:dp2})  around any one of these four fixed points leads to the following set of equations.
\begin{eqnarray}
	\Ddot{\theta}_1 &+& 2 h \beta \dot{\theta}_1 + j(1 - A \sin{\Omega \tau}) \left[ (1 + \mu) \theta_1 - \mu \theta_2  \right] 
	+j \frac{2\beta}{\Omega} A (2\theta_1 - \theta_2) \cos{\Omega \tau }=0,  \label{lin1} \\
	\Ddot{\theta_2} &+& \frac{2 h \beta}{\mu}  \dot{\theta}_2 +l\frac{2\beta}{\mu} \lambda (1-\mu) \dot{\theta}_1  
	+ k\lambda(1 + \mu)(1 - A\sin{\Omega \tau}) (\theta_2-\theta_1) \nonumber\\
	&+&k\frac{2\beta}{\Omega}\frac{\lambda}{\mu}A \left[ (1 + \mu) \theta_2 - 2\mu \theta_1 \right] \cos{\Omega \tau } = 0. \label{lin2}
\end{eqnarray}
where $h = 1$. The integers $j$, $k$, and $l$  can take values either $1$ or $-1$. The set of values,  $h$, $j$, $k$, and $l$,  tabulated below in a row (see Table~\ref{table_1}), corresponds to one of the fixed points of the double pendulum. The coefficients $h$, $j$, $k$, $l$ form a finite group of order $4$, which is identified as Klein-four ($V_4$). The element $h$ serves as the identity. In addition, $j^2 = k^2 = l^2 = h$, $jk=kj=l$, $kl=lk=j$, and $lj = jl = k$. Each row of Table~\ref{table_1} shows an irreducible representation of the Klein-four group corresponding to a fixed point listed in that row.  
\begin{table}[ht]
	\centering
	\begin{tabular}{|c|c|c|c|c|}
		\hline
	
	 Fixed points ($\theta_{1}^*,~\theta_2^*$) & $h$ & $j$ & $k$ & $l$ \\
		\hline
	 $0,0$ & 1 & 1 & 1 & 1\\
		\hline
	 $0,\pi$ & 1 & 1 & -1 & -1\\
		\hline
	$\pi,0$ & 1 & -1 & 1 & -1\\
		\hline
	$\pi,\pi$ & 1 & -1 & -1 & 1\\
		\hline
	\end{tabular}
	\caption{Values of $h$, $j$, $k$ and $l$ for four fixed points ($\theta_{1}^*,~\theta_2^*$) of the coplanar double pendulum.}
	\label{table_1}
\end{table}

The equations of motion for a driven double pendulum get simplified considerably if $\mu = \lambda = 1$. For the fixed point (0, 0) [$h = j = k = l =1$], the equations of motion [Eqs.~(\ref{lin1} and \ref{lin2})] take the following form:
\begin{eqnarray}
	\Ddot{\theta}_1 + 2\beta \dot{\theta}_1 +  \left[ 1 - R \sin{(\Omega \tau - \Delta)} \right] (2\theta_1-\theta_2) &=& 0,  \label{lin1_mu1} \\
	\Ddot{\theta}_2 + 2\beta \dot{\theta}_2 + 2 \left[ 1 - R\sin{(\Omega \tau - \Delta)} \right] (\theta_2 -\theta_1) &=& 0. \label{lin2_mu1}
\end{eqnarray}
The above equations admit solutions where $\theta_2(\tau)= \mp \sqrt{2} \theta_1(\tau)$, that is the angular displacements of the two pendulums are linearly proportional. Equations (\ref{lin1_mu1}) and (\ref{lin2_mu1}) then become identical, and are given as
\begin{equation}
	\ddot{\theta}_{1,2}+2\beta\dot{\theta}_{1,2}+\Omega_{0\pm}^2  \left[1-R \sin{(\Omega \tau - \Delta)} \right] \theta_{1,2}=0, 
	\label{damped_eqn}
\end{equation}
where $\Omega_{0\pm}^2=2\pm \sqrt{2}$. For $A=0$, the solutions $\theta_{j}(\tau)$ for $j=1, 2$ are
\begin{eqnarray}
	&&\theta_2(\tau)=\mp \sqrt{2}\theta_1(\tau)=\exp(-\beta\tau)\nonumber\\
	&&\times\left[\theta_{c\pm}\cos{\sqrt{\left(\Omega_{0\pm}^2-\beta^2\right)}~\tau} +\theta_{s\pm}\sin{\sqrt{\left(\Omega_{0\pm}^2-\beta^2\right)} ~\tau} \right].\nonumber
\end{eqnarray}
The initial conditions fix the constants $\theta_{c\pm}$ and $\theta_{s\pm}$.
   
\subsection{Floquet multipliers}
We now consider the linearized version of the pendulum equations in the absence of damping ($\beta=0$), which simplify (for $h=j=k=l=1$) from Eqs.~(\ref{lin1}) and (\ref{lin2}) as
\begin{eqnarray}
	\Ddot{\theta}_1 +  f(\tau) \left( (1+\mu) \theta_1 -  \mu\theta_2  \right) &=& 0,\label{mathieu1}\\
	\Ddot{\theta}_2 + \lambda (1+\mu) f(\tau) (\theta_2-\theta_1) &=& 0.\label{mathieu2}
\end{eqnarray}
These coupled Mathieu equations are very general and appear in a variety of problems. For example, to understand spatial patterns in a driven Bose-Einstein condensate use a similar model~\cite{Engelss-Atherton_2007,Nguyen_etal_2019}. Optically trapped water droplets may also be excited parametrically. The measured power spectra of position fluctuations of a suspended water droplet were found in excellent agreement with parametrically modulated Langevin dynamics~\cite{Leonardo_etal_2007}. An alternate method of lasing called ``quantum amplification by super-radiant emission of radiation (QASER)" without population inversion could be explained in terms of coupled oscillators with modulated coupling strength and parametric resonance\cite{Svidzinsky_etal_2013,Zhang_etal_2017,Nessler_etal_2020}. The cosmological problem of reheating after inflation could also be thought of as a parametric resonance in the quantum fluctuations of a couple of scalar fields, one heavy and one light~\cite{Kofman_etal_1996}.

Equations (\ref{mathieu1}) and (\ref{mathieu2}) could be decoupled in terms of normal modes, $Y_{\pm}$, given by
\begin{equation}
	\Ddot{Y}_{\pm} + \Omega_{\pm}^2 f(\tau) Y_{\pm}  = 0;~~Y_{\pm}(\tau) = b_{\pm} \theta_1 (\tau) - \theta_2 (\tau),\label{mathieu}
\end{equation}
where 	
\begin{equation}
b_{\pm} =\frac{ (1 - \lambda)(1+\mu) \pm \sqrt{(1+\mu)\left[(\lambda-1)^2 +\mu( \lambda+1)^2\right]}}{2\mu}, \nonumber
\end{equation}
and $\Omega_{\pm}^2 = \mu b_{\pm} + \lambda (1+\mu)$. In Eqs.~(\ref{mathieu}) and henceforth, all terms with upper (lower) signs are taken together. As $f(\tau + T) = f(\tau)$, Eqs. (\ref{mathieu}) are invariant under the transformation $\tau \rightarrow \tau + T$. If $[Y_{1+} (\tau)$, $ Y_{2+} (\tau)]$ and $[Y_{1-} (\tau), Y_{2-} (\tau)]$ are the two sets of independent solutions for the normal modes $Y_{+}(\tau)$ and $Y_{-}(\tau)$, then so are  $[Y_{1+}(\tau +T)$, $Y_{2+}(\tau +T)$] and [$Y_{1-}(\tau +T)$, $Y_{2-}(\tau +T)$]. Therefore, they must transform into linear combinations of themselves as $\tau \rightarrow \tau + T$. It is possible to choose the solutions~\cite{Mechanics_Landau-Lifshitz_1960,Minorsky_1962} such that
\begin{equation}
	Y_{p\pm}(\tau)=\Lambda_{p\pm}^{\tau/T}\Pi_{p\pm}(\tau);~p=1,2.
\end{equation}
In the above,  $\Pi_{p\pm}(\tau+T)=\Pi_{p\pm}(\tau)$ and the Floquet multipliers $\Lambda_{p\pm}$ are constants. As a consequence, one has $Y_{p\pm}(\tau+T)=\Lambda_{p\pm}Y_{p\pm}(\tau)$. It could be shown that the Floquet multipliers obey the following relations for real function $f(\tau)$:
\begin{equation}
\Lambda_{1\pm}\Lambda_{2\pm}=1=\Lambda_{1\pm}^*\Lambda_{2\pm}^*.
\end{equation}
The Floquet multipliers have the following properties:\\
(i) $\Lambda_{1\pm}$ and $\Lambda_{2\pm}$ are real and $\Lambda_{1\pm}=1/\Lambda_{2\pm}$. The solution corresponding to a Floquet multiplier with a magnitude greater than unity will be a growing solution, whereas one with a magnitude smaller than unity will be a decaying solution. \\
(ii) $\Lambda_{1\pm}$ and $\Lambda_{2\pm}$ are complex, $\Lambda_{1\pm}=\Lambda^*_{2\pm}$ with $|\Lambda_{1\pm}|=|\Lambda_{2\pm}|=1$. The solution corresponding to a complex Floquet multiplier will be either a pure oscillatory type or would be executing under-damped oscillations.\\
It is to be noted that the Floquet multipliers $\Lambda_{p+}$
are independent of $\Lambda_{p-}$. So it could happen that one pair of Floquet multipliers is real, and the other is complex. 

Setting $\lambda=1$ in Eqs. (\ref{mathieu1}) and  (\ref{mathieu2}), we get the following  $\mu$ dependent discrete scaling symmetry:
\begin{equation}
	S[\theta_1(\tau)]=\mu \theta_2(\tau),~~ {\rm and}~~S[\theta_2(\tau)]=(1+\mu) \theta_1(\tau), \label{flq1}
\end{equation}
with $S^{-1}$ defined as,
\begin{equation}
	S^{-1}[\theta_1(\tau)]=\frac{1}{1+\mu} \theta_2(\tau),~~ {\rm and}~~S^{-1}[\theta_2(\tau)]=\frac{1}{\mu} \theta_1(\tau). \label{flq2}
\end{equation}
Equations (\ref{mathieu1}) and  (\ref{mathieu2}) remain invariant under the above scale transformation $S$. Equation (\ref{mathieu1}) $\Longleftrightarrow$ Eq.~(\ref{mathieu2}) when the transformation  (\ref{flq1})  is substituted in (\ref{mathieu1}) and  (\ref{mathieu2}).  The transformation  (\ref{flq2}) does the same. This symmetry could be exploited to generate new solutions from a known solution.

\subsection{Floquet analysis}
Floquet analysis is done following the method~\cite{Kumar-Tuckerman_1994} used in the context of Faraday instability. The solutions of Eqs.~(\ref{lin1}) and (\ref{lin2}) may be written, in the Floquet form, as: 
\begin{eqnarray}
	\theta_1(\tau) &=& e^{(s+i{\alpha}\Omega)\tau} \Phi \left(\tau, T=\frac{2\pi}{\Omega}\right) 
	= e^{(s+i{\alpha}\Omega)\tau} \sum_{n=-\infty}^{\infty} \phi_{n} e^{i n\Omega \tau},\label{eq:floquet1} \\ 
	\theta_2(\tau) &=& e^{(s+i{\alpha}\Omega)\tau} \Psi \left(\tau, T=\frac{2\pi}{\Omega}\right) 
	= e^{(s+i{\alpha}\Omega)\tau} \sum_{n=-\infty}^{\infty} \psi_{n} e^{i n\Omega \tau}.\label{eq:floquet2} 
\end{eqnarray}	Here $\Phi$ and $\Psi$ are periodic functions of time with  time period $T= 2\pi/\Omega$, and they are  expanded in Fourier series. The real number $s$ is the growth rate and $\alpha$ ($0 \le \alpha <1$) is the Floquet exponent. The angular displacements $\theta_1$ and $\theta_2$ are real quantities and, therefore, the expansions will be real only if (i) $\alpha = 0$ or (ii) $\alpha = 1/2$. The Fourier modes, which are complex in general, obey the following reality conditions:
\begin{eqnarray}
	\phi_{-n}^{*} &=& \phi_{n} ~~\mbox{and} ~~~
	\psi_{-n}^{*} = \psi_{n} ~~~~~~\mbox{for $\alpha =0$},\nonumber \\
	\phi_{-(n+1)}^{*} &=& \phi_{n} ~~\mbox{and} ~~~
	\psi_{-(n+1)}^{*} = \psi_{n}~\mbox{for $\alpha = 1/2$}.
\end{eqnarray} 
The solutions corresponding to $\alpha = 0$ are called harmonic solutions and they are synchronous to the external driving. Similarly, the solutions corresponding to $\alpha = 1/2$ are called subharmonics and they oscillate with half the driving frequency. Inserting the Floquet expansion of angular displacements [Eqs.~(\ref{eq:floquet1} and \ref{eq:floquet2})] in the linearised equations of motion [Eqs.~(\ref{lin1} and \ref{lin2})] and collecting the coefficients of the $n$th mode in both the equations, we get the following recursion relations:
\begin{widetext}
\begin{eqnarray}
	\mathcal{B}_n\phi_n - j\mu\psi_n &=& A \left[ \left(c\phi_{n-1} + c^* \phi_{n+1}\right) + \left(d\psi_{n-1} + d^*\psi_{n+1}\right) \right] \label{recur1},\\
	\mathcal{C}_n\phi_n + \mathcal{D}_n\psi_n &=& A  \left[  \left(e\phi_{n-1} + e^*\phi_{n+1}\right) +  \left(f\psi_{n-1} + f^*\psi_{n+1}\right)\right], \label{recur2} \\
	\mbox{where}~~\mathcal{B}_n (\beta, \mu, s, \Omega, \alpha) &=& \left[ \{s+i(n+\alpha)\Omega \}^2 + 2 h \beta \{s +i(n+\alpha)\Omega\}+j(1+\mu)\right],\\
	\mathcal{C}_n (\beta, \mu, \lambda, s, \Omega, \alpha) &=& \left[\frac{2l\beta\lambda(1-\mu)}{\mu} \Big\{ s+ i(n+\alpha)\Omega \Big\} - k\lambda (1+\mu)\right],\\
	\mathcal{D}_n (\beta, \mu, \lambda, s, \Omega, \alpha) &=& \left[ \{s+i(n+\alpha)\Omega\}^2+\frac{2h\beta}{\mu}\Big\{s+i(n+\alpha)\Omega \Big\} + k\lambda (1+\mu)\right], \\
	c \left(\frac{\beta}{\Omega}, \mu\right) = &-&j\left[ \frac{2\beta}{\Omega} + i\left(\frac{1+\mu}{2}\right) \right],~
	d \left(\frac{\beta}{\Omega}, \mu \right) =j\left(\frac{\beta}{\Omega}+i\frac{\mu}{2}\right),~ e=-\frac{k}{j}\lambda c,~ f=-\frac{k}{j}\frac{\lambda(1+\mu)}{\mu}d.
\end{eqnarray}
\end{widetext}
The above recursion relations [Eqs.~(\ref{recur1} and \ref{recur2})] may be converted to a generalised  eigenvalue equation:
\begin{equation}
	\boldsymbol{\mathcal{M}} \boldsymbol{\chi} = A \boldsymbol{\mathcal{N}} \boldsymbol{\chi}, \label{geneig}
\end{equation}
\bigskip

\noindent where $\boldsymbol{\chi}$ $=$ $(..., \phi_{-2}, \psi_{-2}, \phi_{-1}, \psi_{-1}, \phi_{0}, \psi_{0}, \phi_{1}, \psi_{1}, \phi_{2}, \psi_{2}...)^{\dagger}$  is a column matrix with $2n$ elements. The square matrices  $\boldsymbol{\mathcal{M}}$ and $\boldsymbol{\mathcal{N}}$ of  size $2n \times 2n$ are given by

\begin{figure*}[!]
	\centering
	\includegraphics[height=!, width=17.0cm] {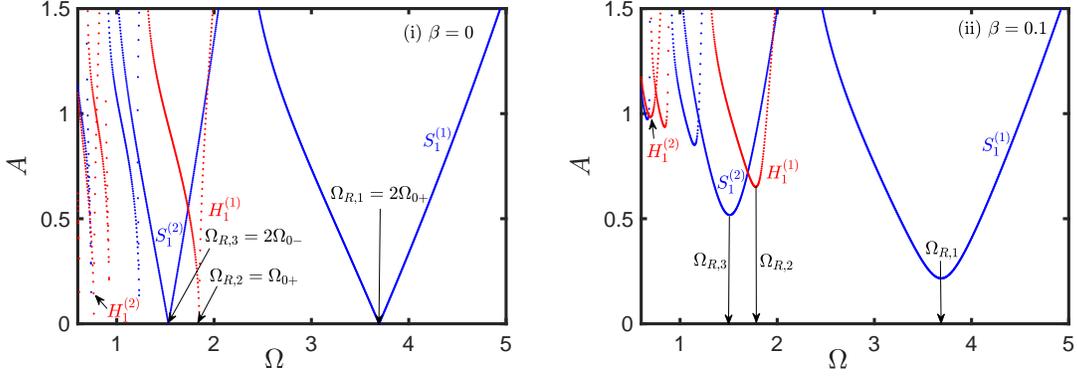}
	\caption{(Colour online) Marginal stability curves for a coplanar double pendulum with length and mass ratios equal to unity ($\lambda = \mu = 1$) in the $\Omega$-$A$ plane for  $\beta = 0$ and $\beta = 0.1$. Blue (black) and red (gray) curves correspond to subharmonic and  harmonic motions, respectively.}
	\label{marginal1}
\end{figure*}

\begin{figure*}[!]
	\centering
	\includegraphics[height=!, width=17.0cm] {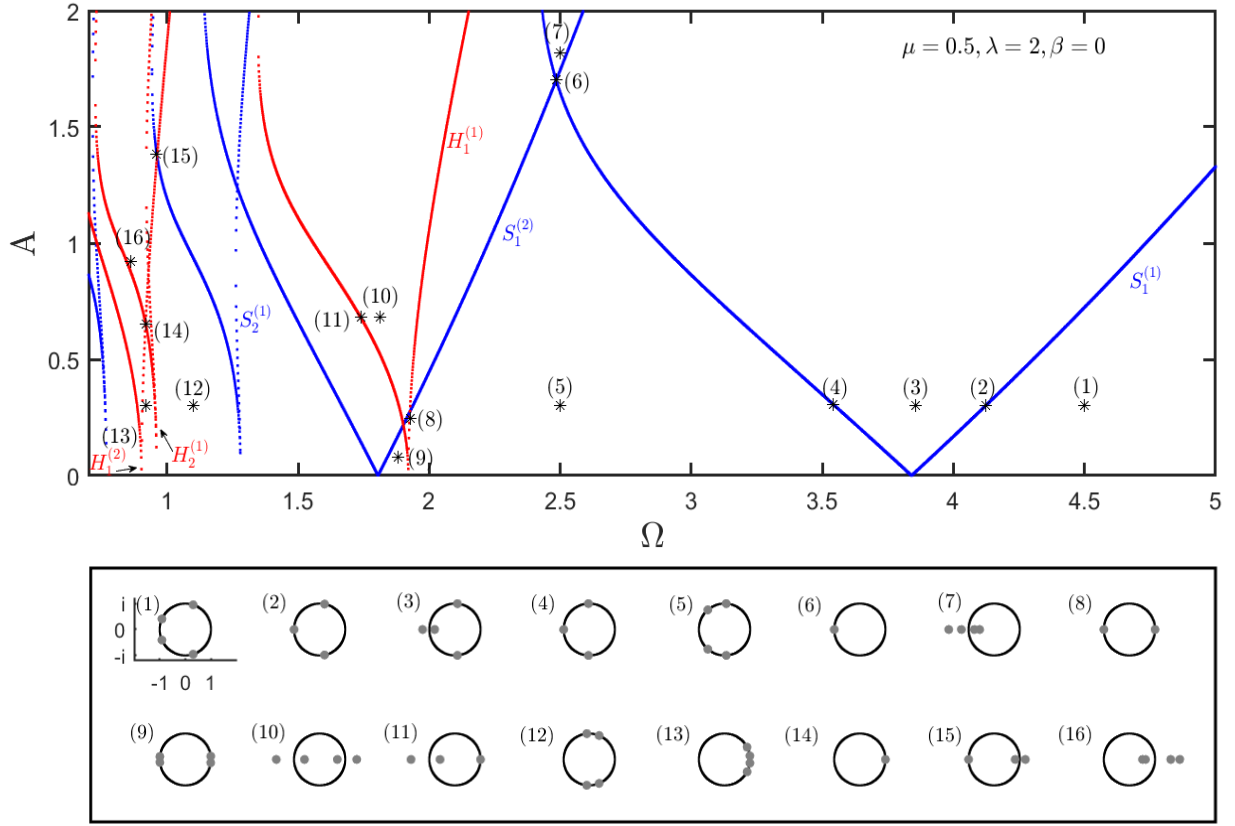}
	\caption{(Colour online) Marginal stability curves for $\lambda=2, \mu= 0.5$ in the $\Omega$-$A$ plane for  $\beta = 0$. Blue (black) and red (gray) curves correspond to subharmonic and  harmonic motions, respectively.}
	\label{fm}
\end{figure*}
 
\begin{widetext}
\begin{equation}
	\boldsymbol{\mathcal{M}} =
	\begin{bmatrix}
	\ddots  & \vdots & \vdots &\vdots  &\vdots &\vdots &\vdots  \\
		\dots &\mathcal{B}_{-1} & -j\mu & 0 & 0  & 0 & 0 &\dots \\
		\dots &\mathcal{C}_{-1}  &  \mathcal{D}_{-1}  &  0 & 0 & 0 & 0 & \dots\\
		\dots & 0 & 0  &\mathcal{B}_0 & -j\mu & 0 & 0 &\dots\\
		\dots & 0 & 0  &\mathcal{C}_0 & \mathcal{D}_0 & 0 & 0 &\dots\\
		\dots & 0 & 0  & 0 & 0 &\mathcal{B}_1 & -j\mu &\dots\\
		\dots & 0 & 0  & 0 & 0 & \mathcal{C}_1 & \mathcal{D}_1 &\dots\\
		&\vdots  &\vdots &\vdots  &\vdots &\vdots &\vdots &\ddots \\
	\end{bmatrix},~~~
	\boldsymbol{\mathcal{N}} =
	\begin{bmatrix}
		\ddots  & \vdots & \vdots & \vdots  & \vdots & \vdots & \vdots \\
		\dots & 0 & 0 & c^* & d^* & 0 & 0 &\dots \\
		\dots & 0 & 0 & e^* & f^* & 0 & 0 &\dots \\
		\dots & c & d & 0 & 0 & c^* & d^* &\dots \\
		\dots & e & f & 0 & 0 & e^* & f^* &\dots \\
		\dots & 0 & 0 & c & d & 0 & 0 &\dots \\
		\dots & 0 & 0 & e & f & 0 & 0 &\dots \\
		&\vdots  &\vdots &\vdots  &\vdots &\vdots &\vdots & \ddots \\
		\end{bmatrix}.
\end{equation}
\end{widetext}
The generalised eigenvalue equation [Eq.~(\ref{geneig})] may be put in the form of a standard eigenvalue equation:
\begin{equation}
	\left(\boldsymbol{\mathcal{N}}^{-1} \boldsymbol{\mathcal{M}}\right) \boldsymbol{\chi} = A \boldsymbol{\chi}~~\mbox{or}~~ \left(\boldsymbol{\mathcal{M}}^{-1} \boldsymbol{\mathcal{N}}\right) \boldsymbol{\chi} = \frac{1}{A}\boldsymbol{\chi}.\label{eigenvalue}
\end{equation}

\begin{figure*}[!]
	\centering
	\includegraphics[height=!, width=17.0cm] {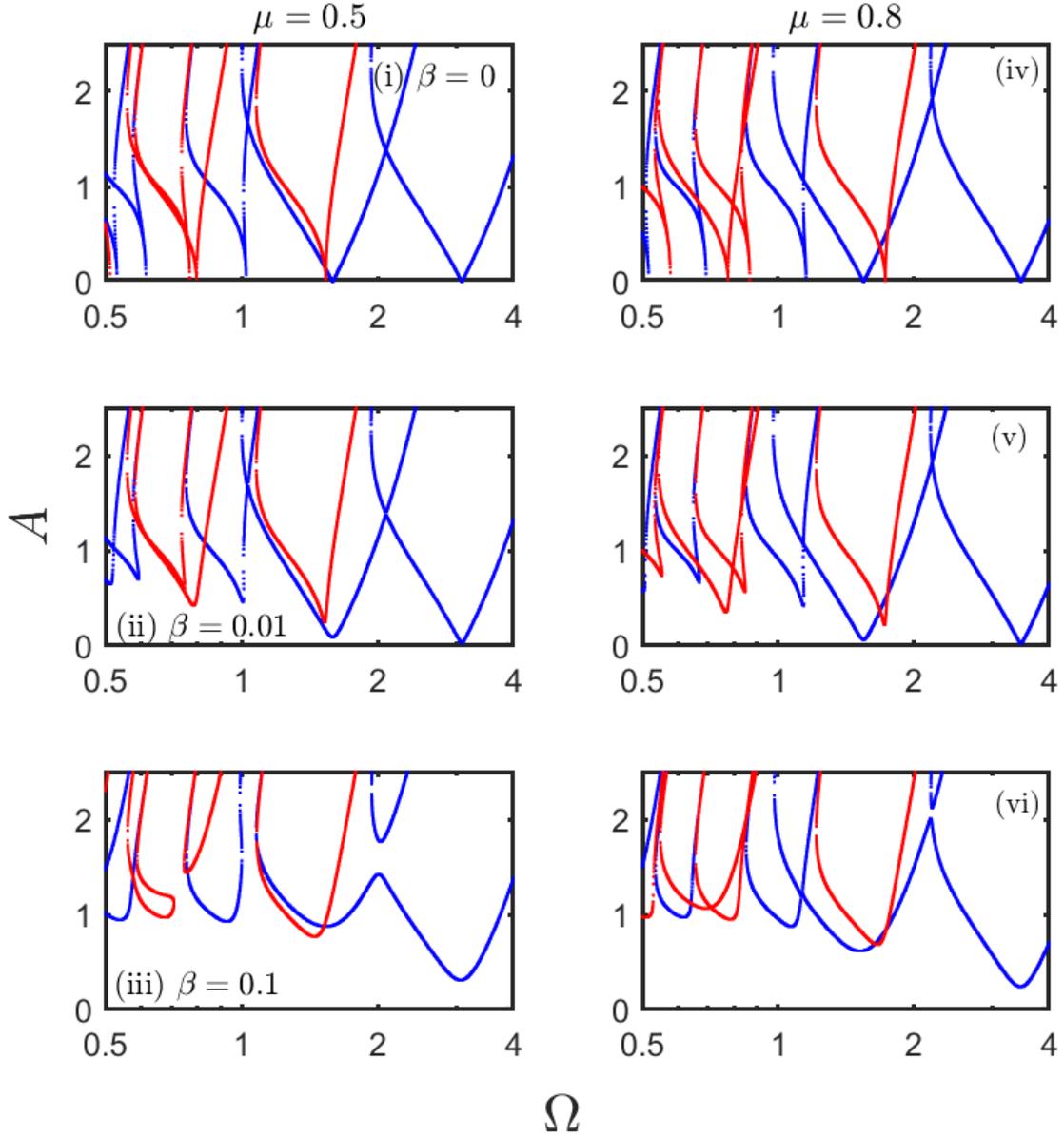}
	\caption{(Colour online) Marginal stability curves for the mass ratio $\mu = 0.5$ (left column) and $\mu = 0.8$ (right column), length ratio $\lambda = 1$ and for different values of damping coefficient: $\beta = 0$ [(i) and (iv)], $0.01$ [(ii) and (v)] and $0.1$ [(iii) and (vi)].   Blue (black) and red (gray) curves are for the Floquet exponent $\alpha = 0$ and $1/2$, respectively.}
	\label{marginal_curves}
\end{figure*}

\begin{figure*}[!]
	\centering
	\includegraphics[height=!, width=17cm] {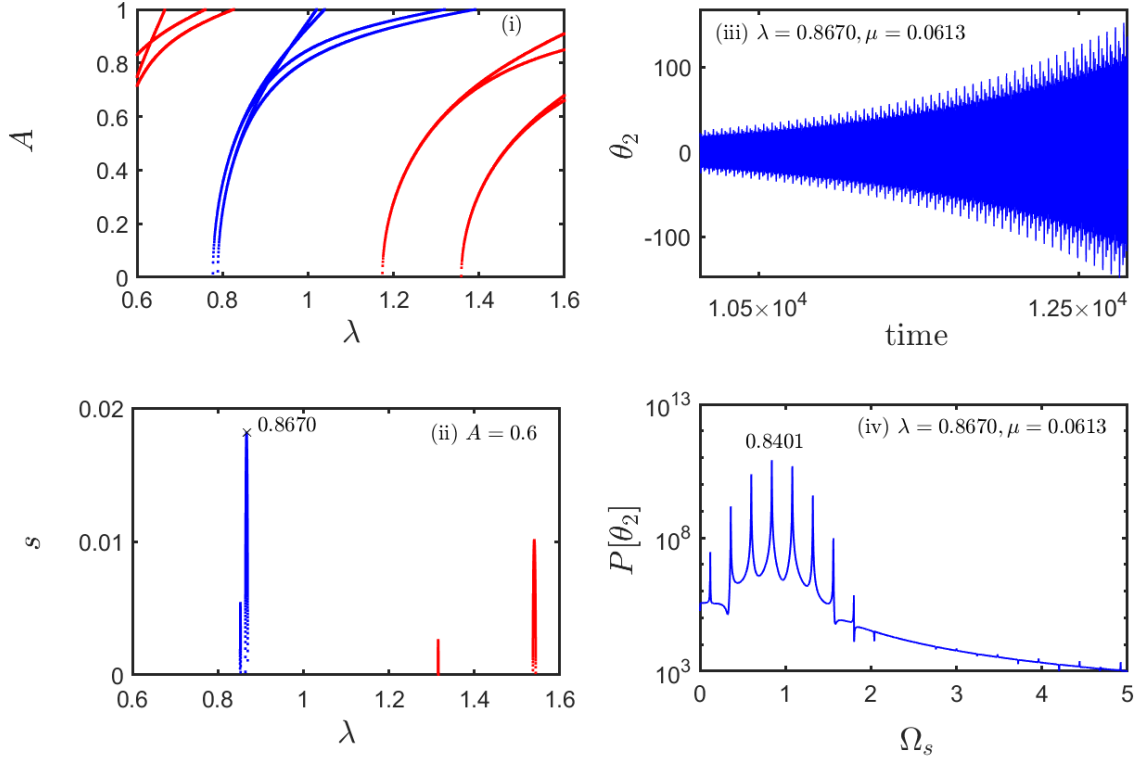}
	\caption{{ (Colour online) 
	   Instability regime (i) and variation of $2\pi s/{\Omega}$ (ii) near higher subharmonics are plotted as a function of $\lambda$ for a coplanar double pendulum for a fixed driving frequency $\Omega = 0.24$ and driving amplitude $A = 0.6$. The solution (iii) and its power spectrum density (iv) are also shown at the largest Floquet exponent at $\lambda = 0.8670$, $\mu = 0.0613$, $\beta = 0.0$.}} \label{combres}
\end{figure*}

\begin{figure*}[!]
	\centering
	\includegraphics[height=!, width=17cm] {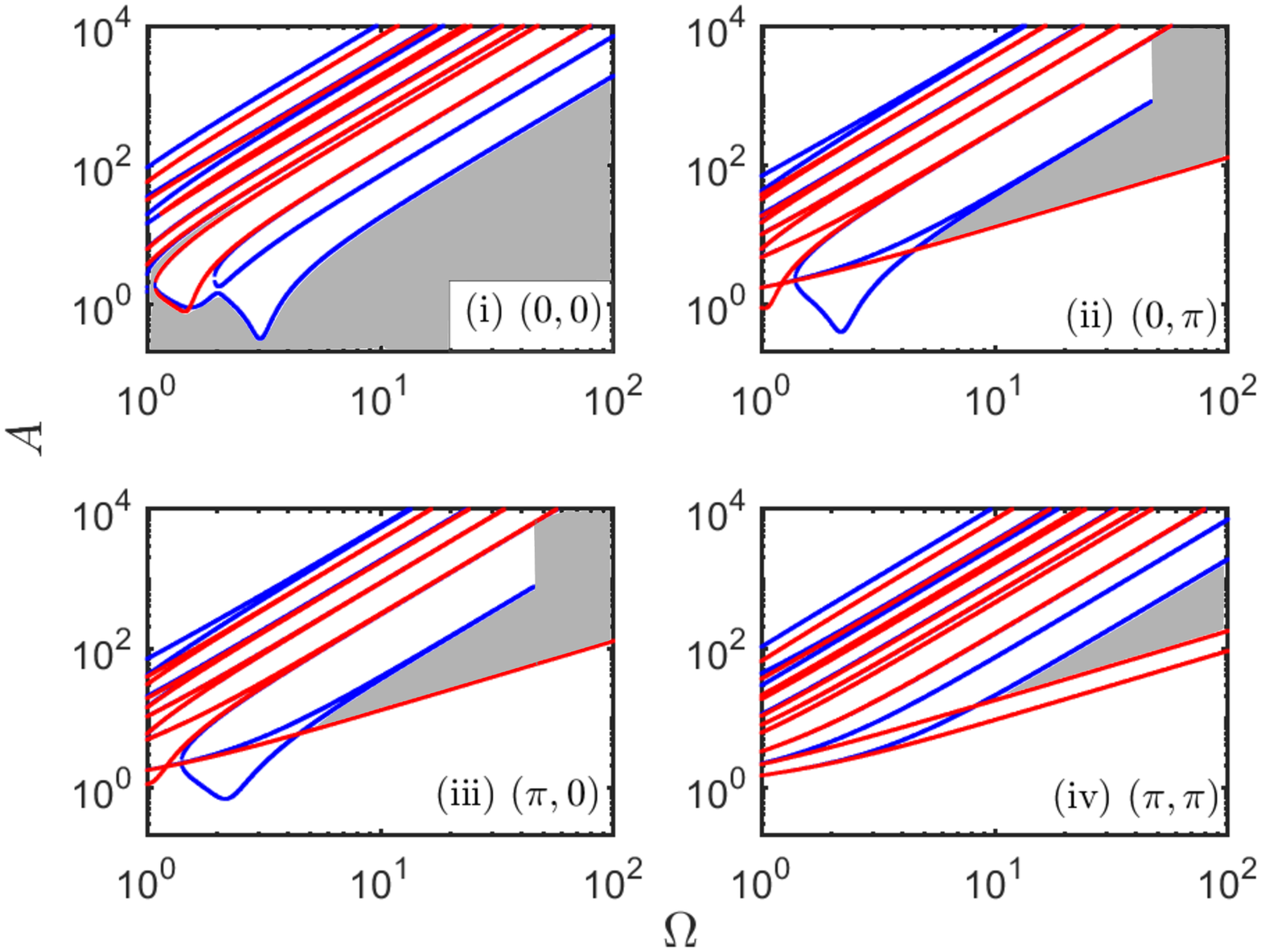}
	\caption{(Colour online) Marginal stability curves for fixed points (i) ($0$, $0$), (ii) ($0$, $\pi$), (iii) ($\pi$, $0$), (iv) ($\pi$, $\pi$) in the $\Omega$-$A$ plane for $\mu = 0.5$, $\lambda = 1$ and  $\beta = 0.1$.
	The shaded area in a viewgraph is the region of stability for the fixed point mentioned in that viewgraph against periodic perturbations.}
	\label{Inverted_double_stability_mu0,5_combined}
\end{figure*} 	
	
As $n$ varies from $-\infty$ to $+\infty$, the matrix is of infinite size. In practice one has to truncate the matrix to a suitable finite size and find eigenvalues of the matrix. The real and positive eigenvalues of Eq.~(\ref{eigenvalue}) give possible values the dimensionless amplitude $A$ or its inverse ($1/A$). By varying the driving frequency $\Omega$ in small steps the eigenvalues $A(\Omega)$ or $1/A(\Omega)$ can be computed for each value of $\Omega$ and fixed values of parameters $\beta, \mu, \lambda, s, \alpha$. The eigenvalues corresponding to the growth rate $s = 0$ fall on the neutral (marginal) stability curves for these values of the parameters. The eigenvalues of the square matrix  $\boldsymbol{\mathcal{N}}^{-1} \boldsymbol{\mathcal{M}}$ or the matrix $\boldsymbol{\mathcal{M}}^{-1} \boldsymbol{\mathcal{N}}$ are found to be either pairs of complex conjugate numbers or real numbers. As the driving amplitude $A$ is real, only real and positive eigenvalues are relevant. Their number depends on the size of square matrix $\boldsymbol{\mathcal{N}}^{-1} \boldsymbol{\mathcal{M}}$ or $\boldsymbol{\mathcal{M}}^{-1} \boldsymbol{\mathcal{N}}$.

\begin{figure*}[!]
\centering
 \includegraphics[height=!, width=17.0cm] {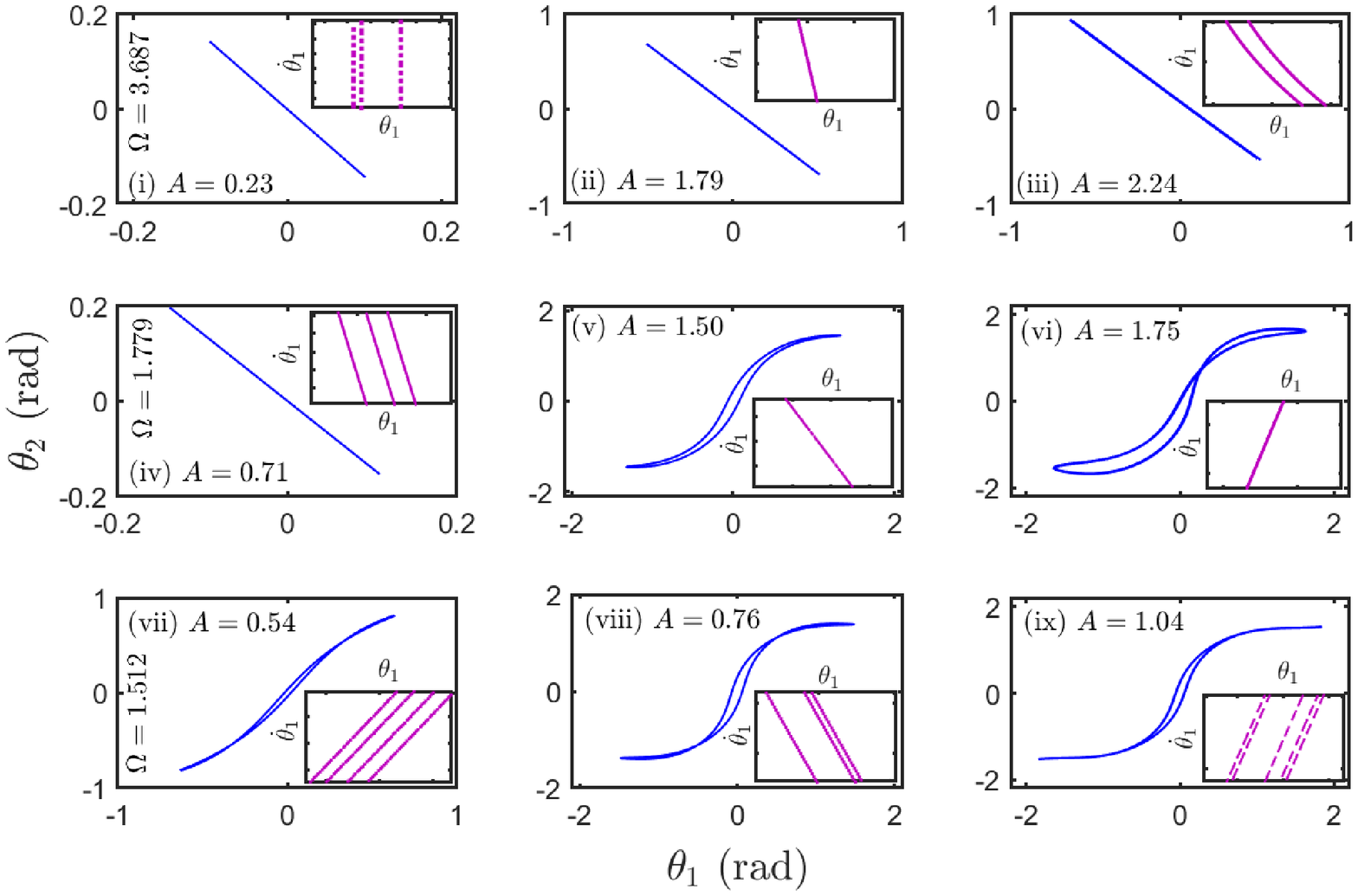}
 \caption{(Colour online) 
Phase portraits in the configuration space [the $\theta_1$-$\theta_2$ ($X_1$-$X_3$) plane] for a coplanar double pendulum ($\mu = \lambda = 1$, $\beta = 0.1$) when the its point of support is vibrated sinusoidally in the vertical direction. Three different values of dimensionless driving amplitude $A$  are considered for each of three values of the driving frequency $\Omega = 3.687$ [(i)-(iii)], $1.779$ [(iv)-(vi)] and  $1.512$ [(vii)-(ix)]. Inset in each plot displays a zoomed view of a tiny portion of the corresponding limit cycle in the $\theta_1$ - $\dot{\theta}_1$ plane [magenta (gray) curves].} \label{PhasePlot1}
\end{figure*}

\section{Results of the Floquet analysis}

 We have considered the length ratio, $\lambda$ between $1/2$ and $2$ in this article. The stability matrix $\boldsymbol{\mathcal{M}}^{-1} \boldsymbol{\mathcal{N}}$ (or $\boldsymbol{\mathcal{N}}^{-1} \boldsymbol{\mathcal{M}}$) is truncated to a reasonable size so that the first few lower eigenvalues $A$ (real and positive ones) are determined with preassigned accuracy for each value of $\Omega$. All the results reported here are for the expansion of $\theta_1$ and $\theta_2$ up to ten Fourier modes. This means that a matrix of size $40 \times 40$ for $\alpha = 1/2$ and $42 \times 42$ for $\alpha = 0$ with a preassigned accuracy of $10^{-6}$. The growth rate $s$ is set equal to zero, and small values the damping coefficient $\beta$ are considered. The real positive eigenvalues are recorded for a given value of the dimensionless driving frequency $\Omega$. The process is repeated by varying $\Omega$ in small steps of $\Delta \Omega = 0.003$. The eigenvalues are plotted for each value of $\Omega$ in the $\Omega$-$A$ plane for $\alpha = 1/2$ and $\alpha = 0$. These eigenvalues form tongue-like regions in the parameter space (the $\Omega$-$A$ plane), and separate the regions of stability ($s < 0$) from those of instability ($s > 0$).

\subsection{Stability of a normal double pendulum} 

We first present the marginal stability curves computed for a parametrically driven coplanar double pendulum with two equal masses and equal lengths ($\mu=\lambda=1$). Figure \ref{marginal1} shows the marginal stability curves for (i) $\beta = 0$ and (ii) $\beta = 0.1$. The regions inside blue (black) curves are subharmonic instability zones ($\alpha = 1/2$), while those inside red (gray) curves are  harmonic instability zones ($\alpha =0$). Regions marked by symbols $S_1^{(1)}$ and $S_1^{(2)}$ ($H_1^{(1)}$ and $H_1^{(2)}$) in the $\Omega$-$A$ plane stand for the first subharmonic (harmonic) instability zones corresponding to the normal modes with frequencies $\Omega_{0+}$ and $\Omega_{0-}$, respectively. The frequencies corresponding to the minima of different instability zones are resonance frequencies. For $\beta = 0$, the largest frequencies at subharmonic (harmonic) resonances are $2\Omega_{0+}$ ($\Omega_{0+}$) and $2\Omega_{0-}$ ($\Omega_{0-}$). In the presence of damping ($\beta = 0.1$) as shown in Fig.~\ref{marginal1} (ii),  the marginal boundaries move away from the frequency axis. The  driving amplitude to excite the double pendulum subharmonically at the highest resonance frequency is the least.  This results in an out-of-phase motion. 
  
The marginal stability curves  computed for the case of two unequal masses ($m_1 \neq m_2$) for the fixed point $(\theta_1^*,  \theta_2^*) = (0, 0)$ are now discussed.  The elements of the Klein-four group for this case are: $h=j=k=l=1$. The upper panel of Fig.~\ref{fm} shows the marginal stability curves for $\lambda=2, \mu= 0.5$ in the absence of damping. The lower panel shows the Floquet multipliers corresponding to a variety of points numbered in the $\Omega$-$A$ plane. All the four Floquet multipliers for the points outside the instability zones (see, points $1, 5, 9, 12, 13$) are complex. They form two sets of complex conjugate pairs. The magnitude of real parts of these multipliers are less than unity and they represent decaying periodic solutions. When the points are chosen on the marginal stability curves, then at least one pair of the  multipliers is real. If the point falls on the subharmonic (harmonic) marginal curve, then at least a pair of multipliers will be $-1$ ($+1$) as shown for the points $2, 4, 6, 8, 14$. If the point is inside only one of the instability zones, then at least one pair of multipliers will be real. If the point is inside a subharmonic (harmonic) instability zone, then the real multipliers will be negative (positive), as shown for the point $3$. If the point is inside two instability zones, then all the four multipliers are real (see, points $7, 10, 16$).  The point $11$ ($15$)  is on the boundary of a harmonic (subharmonic) instability zone and just inside a subharmonic (harmonic) zone.

Figure \ref{marginal_curves} displays marginal stability curves for the stable fixed point $(0, 0)$  for mass ratio,  $\mu = 0.5$ (the left column) and $\mu = 0.8$ (the right column) for different values of $\beta$. The regions inside blue (black) curves are the zones of subharmonic instability ($\alpha = 1/2$), while those inside red (gray) curves are zones of harmonic instability ($\alpha = 0$). For any point outside all these instability zones in the $\Omega$-$A$ plane, the double pendulum will not have any periodic solutions. The marginal curves plotted in the top, middle and bottom rows are for $\beta = 0$, $0.01$, and $0.1$, respectively.  The instability zones for $\alpha = 1/2$ never cross those for $\alpha = 0$ for the same normal mode, as they belong to two different classes of oscillatory solutions. However, the marginal stability curves corresponding to two different normal modes may have a common point. As there are several instability zones, there are several resonance frequencies. All the marginal curves touch the frequency axis [Figs.~\ref{marginal_curves} (i), (iv)] in the absence of damping ($\beta = 0$). The double pendulum may be parametrically excited at an infinitesimal driving amplitude in this case.

The damping induces some novel behaviour in the form of merging of subharmonic tongues for both $\mu=0.5$ and $\mu=0.8$. The marginal curves for $\beta = 0.01$ are shown in Figs.~\ref{marginal_curves} (ii) and \ref{marginal_curves} (v). When the damping is further increased ($\beta = 0.1$) some drastic reorganization of merged zones are recorded for $\mu=0.5$. The stability zones for $\mu = 0.5$ and $\mu = 0.8$ look visibly different for $\beta = 0.1$ in Fig.~\ref{marginal_curves} (iii)) and ~\ref{marginal_curves} (vi). This indicates that the dynamical behaviour of a double pendulum with the same driving frequency and  amplitude may differ significantly for different mass ratios ($\mu$). A relatively smaller driving amplitude is capable of exciting it with larger value of $\mu$ at the largest subharmonic resonance frequency for a fixed value of $\beta$. In the frequency range ($1 < \Omega < 1.2$) shown in Fig.~\ref{marginal_curves}(iii), the reorganization of merged subharmonic zones introduces gaps in the $\Omega$-$A$ plane where no effort can excite a periodic oscillation about the fixed point $(0, 0)$. For $\mu=0.8$  [see Fig.~\ref{marginal_curves} (vi)], we find no such window devoid of periodic motion in the $\Omega$-$A$ plane. The linear stability analysis provides the ground for initial guesses about the dynamics of parametrically driven double pendulum. However, the actual dynamics is governed by the full set of equations [Eqs.~(\ref{eq:dp1}) and (\ref{eq:dp2})]. Therefore, nonlinear analysis of the system is required for thorough understanding of its dynamics.

We now fix the driving frequency $\Omega=0.24$ and vary the one of the pendulum parameters in the absence of damping ($\beta= 0$). This is equivalent to fixing the driving frequency and choosing different double pendulums, as the normal mode frequencies depend on $\lambda$ and $\mu$. It is similar to the process followed in the model of a QASER~\cite{Svidzinsky_etal_2013,Zhang_etal_2017,Nessler_etal_2020}, where they reported a large frequency response (gain) for a small forcing frequency.  We have used a relation: $1+\mu = 64\lambda/[15 (\lambda + 1)^2] $ which sets $3\Omega_{+}^2 = 5\Omega_{-}^2$. The difference $\Delta \Omega = \Omega_{+} (\lambda)-\Omega_{-}(\lambda)$ becomes a function of $\lambda$ only.  Figure \ref{combres} (i) shows the instability regimes in the $A-\lambda$ plane for a fixed value $\Omega$ and $\beta=0$. The onset of subharmonic instabilities ($A\rightarrow 0$) are at $\lambda=0.7775$ and $0.7889$. The former corresponds to $\Omega = 2 \Omega_{+}(0.7775)/9$ and the latter corresponds to $\Omega = 2 \Omega_{-}(0.7889)/7$. Similarly, harmonic instabilities at the onset are found at $\lambda=1.174$ and $1.359$. They correspond to $\Omega=2 \Omega_{+}(1.174)/10$ and $\Omega=2 \Omega_{-}(1.359)/8$, respectively. For higher values of $A$, the marginal curves are titled towards higher values of $\lambda$.  We observe closely placed narrow subharmonic instability zones [blue (black) curves] and two harmonic instability zones [red (gray) curves]. These regimes become relatively wider at the higher values of $A$. Inside this instability zones, the real part of the Floquet exponent is positive. Figure \ref{combres} (ii) displays  $2 \pi s/{\Omega}$ as a function of $\lambda$  for $A=0.6$. The closely placed blue (black) peaks at lower $\lambda$ are for the subharmonic instability zones, while the two red (gray) peaks at higher $\lambda$ are for the harmonic instability zones. The maximum of the  largest peak is located at $\lambda = 0.8670$, and the corresponding value of $\mu = 0.0613$. At $\lambda= 0.8670$ the $\Delta \Omega \approx \Omega$. This has resemblance with the results of the QASER model~\cite{Zhang_etal_2017}, although our system is different. Figure \ref{combres} (iii) shows the time evolution of $\theta_2$ after the transients die out. As the $\theta_2$ grows with time, peaks in its power spectrum density (PSD) will keep growing with time. However, the positions of the peaks in its PSD [Fig.~\ref{combres} (iv)] are located at the same frequencies. The maximum power of the growing solution is concentrated at frequency $\Omega_s = 0.8401$, which is equal to $7\Omega/2$. In this case, the dominating response frequencies are around four times the driving frequency.

\subsection{Partially or fully inverted double pendulum}
We now discuss the results of linear stability of a double pendulum, in which at least one of the  two pendulums is inverted.  Figure~\ref{Inverted_double_stability_mu0,5_combined} displays the marginal stability curves obtained using the Floquet method for different equilibrium points in the $\Omega$-$A$ plane for $\mu = 0.5$ and $\beta = 0.1$.  The blue (black) and red (gray) curves ($s = 0$) are for subharmonic and harmonic instabilities, respectively. The curves shown in Fig.~\ref{Inverted_double_stability_mu0,5_combined} (i) are stability boundaries of a normal double pendulum, which corresponds to the fixed point $\theta_1^* = \theta_2^* = 0$. The shaded zone is the region of parameter space for which the fixed point $(0, 0)$ is stable against periodic perturbations.   It  becomes unstable as the driving amplitude and frequency are within any of the instability zones. The shaded regions in Figs.~\ref{Inverted_double_stability_mu0,5_combined} (ii) and (iii) are the zones of stability of fixed points $\theta_1^* = 0, \theta_2^* = \pi$ and $\theta_1^* = \pi, \theta_2^* = 0$, respectively. The stability boundaries for fixed points  $(0, \pi)$ and $(\pi, 0)$ closely resemble each other and become identical  for a double pendulum of equal masses ($\mu = 1$).
In this case the boundary for the harmonic instability [red (gray) curves] is below those for subhamonic solutions [blue (black) curves] for $\Omega$ greater than certain value $\Omega^{*}$. The values of $\Omega^{*}$ for the fixed points ($0, \pi$) and ($\pi, 0$) are $4.521$ and $4.481$, respectively.  For the fixed point ($0, \pi$), the lowest curve for the subharmonic instability [blue (black) curve] terminates at $\Omega = 47.25$ and $A = 827.7$ for $\beta  = 0.1$.  If $\beta = 0$, then this curve does not terminate (not shown here) and continues. The analogous termination point for the fixed point ($\pi, 0$) is found at $\Omega = 45.62$ and $A = 771.6$ for the same $\beta$. It is observed that for both the cases of partially inverted pendulums, $A/\Omega^2 = a/l_1 = 0.3707$ at the termination point. The damping enlarges the region of stability of a partially inverted double pendulum in the parameter space. Figure \ref{Inverted_double_stability_mu0,5_combined} (iv) shows the stability boundaries for the fully inverted double pendulum, which corresponds to the fixed point $\theta_1^* = \theta_2^* = \pi$. In this case, two harmonic instability boundaries are below that for the subharmonic instability for  $\Omega > \Omega^{*} = 8.231$. The shaded region in the parameter space is the region of stability of a fully inverted double pendulum. This region is consistent with the pendulum theorem~\cite{Acheson_1993}, although the limiting boundaries given in Ref.~\cite{Acheson_1993} lie inside and very close to the upper and the lower boundaries of the shaded region shown in Fig.~\ref{Inverted_double_stability_mu0,5_combined} (iv).  In addition,  the stability zones at low driving amplitudes for different fixed points overlap. This may lead to a possibility of more than one stable fixed point depending on the initial conditions. At higher frequencies and lower amplitudes of driving, the double pendulum may be maintained in an either fully inverted state ($\pi, \pi$) or in  one of the partially inverted states [($0, \pi$) and ($\pi, 0$)]. At larger values of $A$, the actual behaviour of the system will be governed by the full nonlinear equations.

\begin{figure*}[!]
	\centering
	\includegraphics[height=!, width=17.0cm] {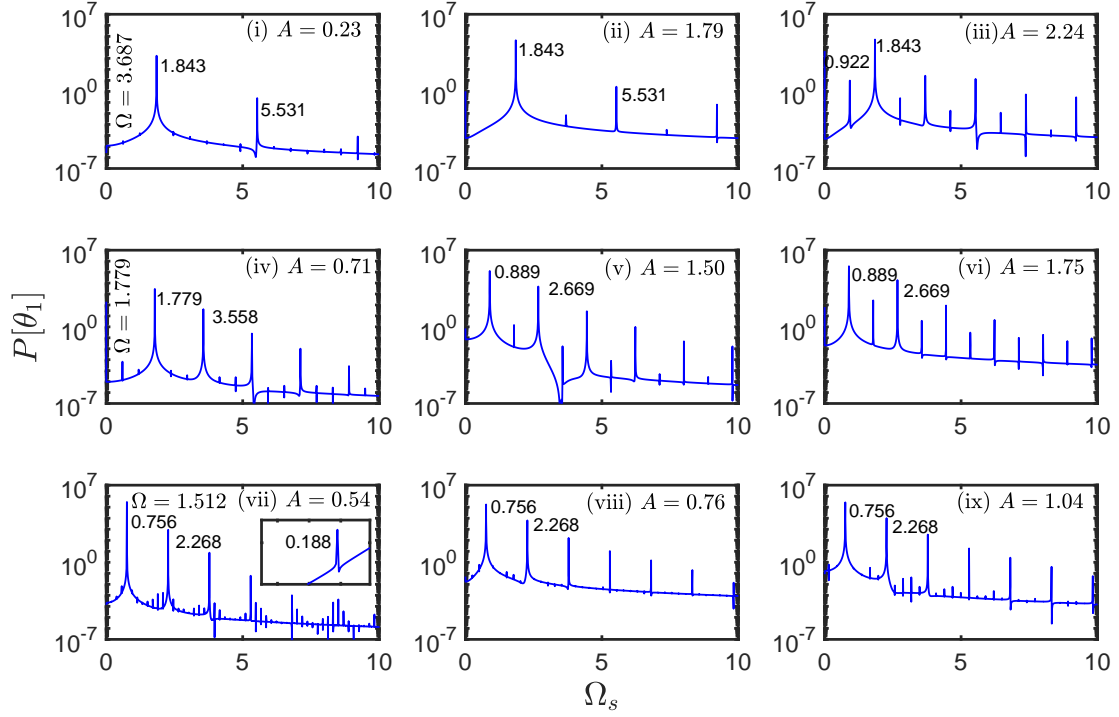}
	\caption{(Colour online) Power spectra of the time signal for the angular displacement $\theta_1 (\tau)$ when the point of support of the double pendulum is vibrated vertically. All the parameters are as given in Fig.~\ref{PhasePlot1}. Inset of plot (vii) is an enlarged view of the first peak in the power spectrum. } \label{PS1}
\end{figure*}

\begin{figure*}[!]
	\centering
	\includegraphics[height=!, width=16.0cm] {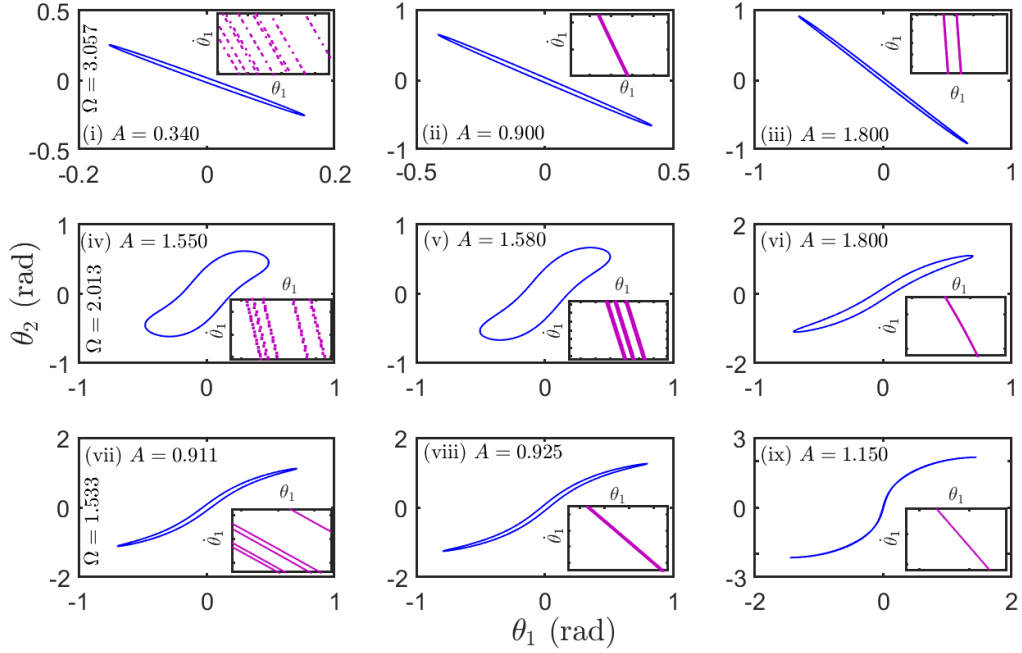}
	\caption{(Colour online) Phase portraits in the configuration space [$\theta_1$-$\theta_2$ ($X_1$-$X_3$) plane] for a coplanar double pendulum with its pivot vibrated vertically. Parameters are: $\mu = 1/2$, $\lambda = 1$ and $\beta = 0.1$. Different values of dimensionless driving amplitude $A$ are considered for $\Omega = 3.057$ [(i)-(iii)], $2.013$ [(iv)-(vi)] and  $1.533$ [(vii)-(ix)]. Inset of each viewgraph shows a tiny part of the zoomed view of the corresponding limit cycle in the $\theta_1$ - $\dot{\theta}_1$ plane [magenta (gray) curves].} \label{PhasePlot2}
\end{figure*} 

\begin{figure}[!]
	\centering
	\includegraphics[height=!, width=8.0cm] {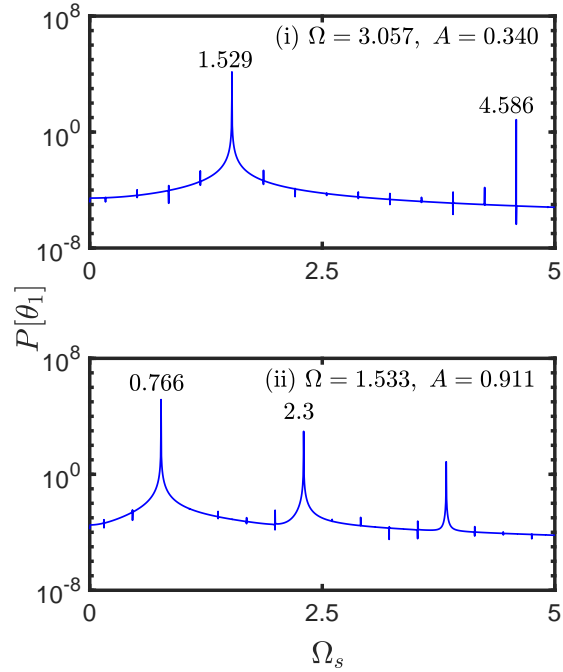}
	\caption{(Colour online) Power spectra of the time signal for the angular displacement $\theta_1 (\tau)$. The upper (lower) plot corresponds to the limit cycles shown in Fig.~\ref{PhasePlot2} (i) [\ref{PhasePlot2} (vii))] when the point of support is vibrated vertically.}
	\label{PS2}
\end{figure}

\begin{figure*}[!]
	\centering
	\includegraphics[height=!, width=16.0cm] {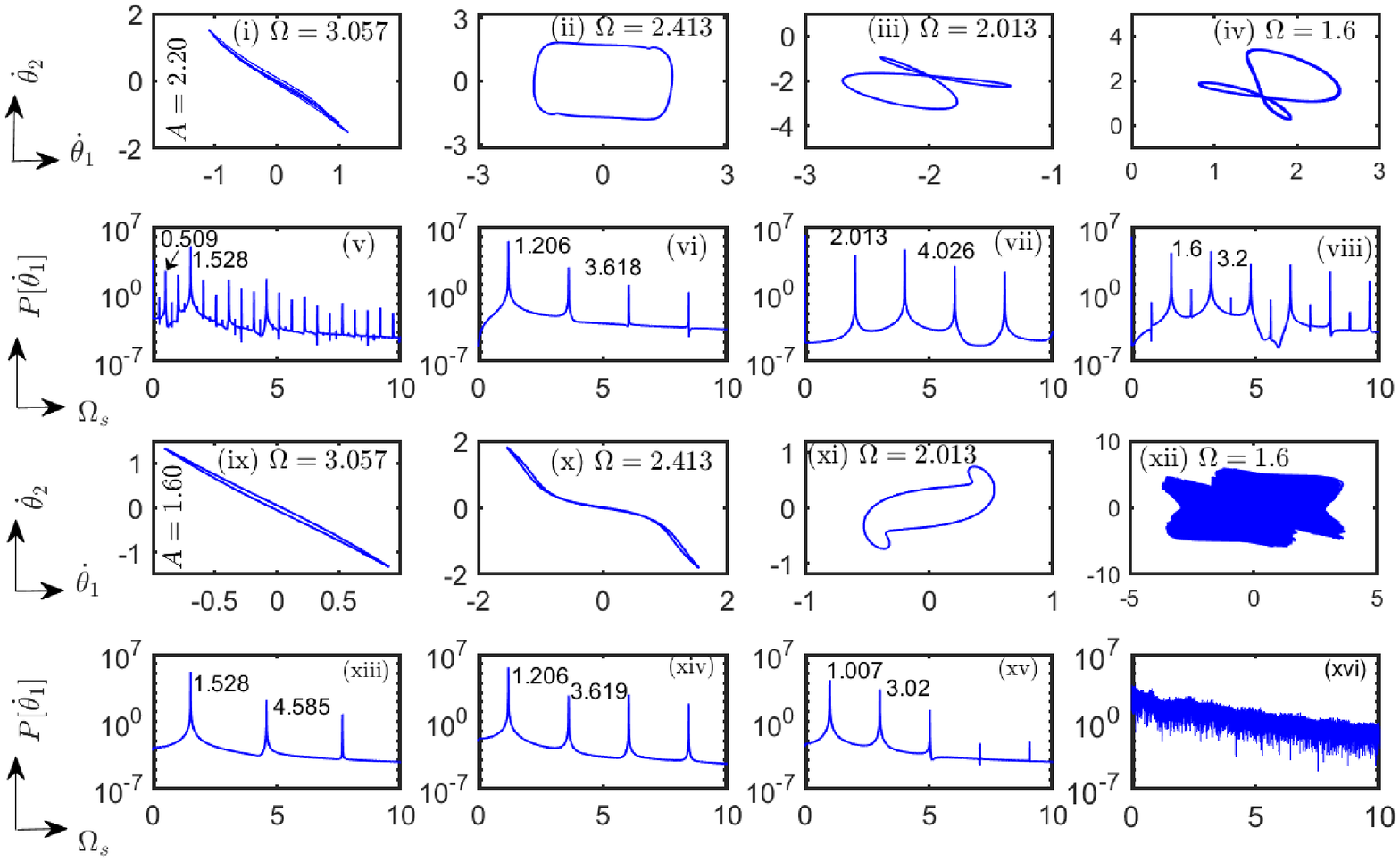}
	\caption{(Colour online) Large amplitude  oscillations and rotational motion of a coplanar double pendulum with its pivot vibrated vertically. Phase portraits in the plane of generalized velocities [$\dot{\theta}_1$-$\dot{\theta}_2$ ($X_2$-$X_4$) plane] and power spectra for the corresponding angular velocity $\dot{\theta}_1$ for different values of $\Omega$ for a fixed value of $A$. Parameters are: $\mu = 1/2$, $\lambda = 1$ and $\beta = 0.1$.} \label{PhasePlot3}
\end{figure*} 

\section{Nonlinear motion under parametric driving}

The phenomenon of parametric resonance occurs in nonlinear optical systems described by a nonlinear Schr\"{o}dinger equation~\cite{Ripoll_Gracia_1999}. They also occur in an extended system like a taut string with periodically modulated tension~\cite{Rowland_2004}.  These system could be described by modified Mathieu equation with nonlinearity and damping. The nonlinear motion of a double pendulum involves coupled Mathieu equations with nonlinearilty and damping which are more general in nature. To investigate the nonlinear dynamics of a parametrically driven damped coplanar double pendulum, the equations of motion may be converted to a dynamical system, which may be integrated numerically.
For this purpose, we introduce dynamical variables as $X_1 = \theta_1$, $X_2 = \dot{\theta}_1$, $X_3 = \theta_2$, $X_4 = \dot{\theta}_2$ and $X_5 = \Omega \tau$ and rewrite the non-linear equations  (\ref{eq:dp1} and \ref{eq:dp2}) as: 
\begin{widetext}
	\begin{eqnarray}
		\dot{X_1} &=& X_2, \label{dynsys1}\\ 
		\dot{X}_2 &=& -\frac{1}{\left[ 1+\mu\sin^2{(X_1-X_3)}\right]} \Big[ 
		(1-A\sin{X_5}) \Big\{ (1+\mu)\sin{X_1}-\mu\cos{(X_1-X_3)}\sin{X_3}\Big\} \nonumber \\ 
		&+&	2\beta\left\{1+\sin^2{(X_1-X_3)}\right\}X_2   +\frac{2\beta}{\Omega}A\Big\{ 2\sin{X_1}-\cos{(X_1-X_3)}\sin{X_3} \Big\} \cos{X_5} \nonumber \\
		&+&\frac{\mu}{\lambda}\left\{ X_4^2 + \lambda X_2^2 \cos{(X_1-X_3)}\right\}\sin{(X_1-X_3)} \Big], \label{dynsys2}\\
		\dot{X}_3 &=& X_4, \label{dynsys3}\\
		\dot{X}_4 &=& -\frac{1}{\left[ 1+\mu\sin^2{(X_1-X_3)}\right]} \Big[ \Big\{ \lambda(1+\mu)(1-A\sin{X_5}) \left[ \sin{X_3}- \cos{(X_1-X_3)} \sin{X_1}\right] \Big\} \nonumber \\ 
		&+&\frac{2\beta}{\mu}\Big\{ \left( 1 + \mu \sin^2{(X_1-X_3)}\right) X_4 +\lambda(1-\mu)X_2\cos{(X_1-X_3)} \Big\} \nonumber \\
		&+&\frac{2\beta\lambda}{\Omega\mu}A\Big\{ (1+\mu) \sin{X_3} - 2\mu\cos{(X_1-X_3)} \sin{X_1} \Big\}\cos{X_5} \nonumber \\
		&-& \Big\{ \lambda (1+\mu)X_2^2 + \mu X_4^2 \cos{(X_1 - X_3)} \Big\} \sin{(X_1 - X_3)} \Big],\label{dynsys4}\\
		\dot{X}_5 &=& \Omega.\label{dynsys5}
	\end{eqnarray}
\end{widetext}

The above dynamical system describes the motion of a coplanar double pendulum whose point of support is vibrated sinusoidally in the vertical direction. As the equations of motion are nonautonomous, the phase space is five-dimensional. However, the variable $X_5$ keeps growing linearly with time but the term $\sin{X_5}$ remains bounded for all time. Interesting dynamics is effectively in the four dimensional phase space of variables $X_1$, $X_2$, $X_3$ and $X_4$.   The dynamical system is integrated numerically using a standard fourth-order Runge-Kutta (RK4) method for given values of all the parameters ($A$, $\Omega$, $\lambda$, $\mu$ and $\beta$). The dimensionless time steps $\Delta \tau$ are chosen equal to $0.005$ for all the phase portraits shown here. Initial values of the dynamical variables $X_i$ ($i = 1, 2, 3, 4$) are chosen randomly in a small neighbourhood of each of the four fixed points. The initial value of $X_5$ is always set equal to zero. All the results reported here for oscillatory swings of a double pendulum are found to be independent of the choice of initial conditions.

\subsection{Excitation of a normal double pendulum}
		
Figure~\ref{PhasePlot1} shows oscillatory motion of a driven double pendulum having equal masses and with $\beta = 0.1$. The projection of corresponding limit cycles on the configuration space [$\theta_1$-$\theta_2$ (or, $X_1$-$X_3$) plane] is displayed for different values of the driving frequency $\Omega$ and amplitude $A$. Inset of each plot of Fig.~\ref{PhasePlot1} shows a zoomed view of a small part of the projection of the corresponding limit cycle on the $\theta_1$-$\dot{\theta}_1$ plane [magenta (gray) curves] to clearly display the details of orbits in the limit cycle.  The power spectra of the phase variable $\theta_1$ for all the plots [Fig.~\ref{PhasePlot1} (i)-(ix)] are displayed in Fig.~\ref{PS1} (i)-(ix). The upper row [Fig.~\ref{PhasePlot1} (i)-(iii)] is for $\Omega = 3.687$, which is very close to the first resonance frequency $\Omega_{R,1}$ for $\beta = 0.1$ [see Fig.~\ref{marginal1} (ii)]. Figure~\ref{PhasePlot1} (i) shows an out-of-phase oscillatory motion for the driving amplitude $A = 0.23$, which is just above the onset of $S_1^{(1)}$. The limit cycle is squeezed to almost a straight line in the configuration as well as in the velocity spaces. This occurs as $\theta_2 = b \theta_1$ with $b \approx -\sqrt{2}$, as predicted by the linear theory [Eqs.~(\ref{lin1_mu1} and \ref{lin2_mu1})]. However, a closer look at the limit cycle reveals nonlinear effects. There are three closely spaced orbits in the phase space, as shown in the inset of Fig.~\ref{PhasePlot1} (i). The power spectrum of $\theta_1$ [Fig.~\ref{PS1} (i)] has the largest peak at the response frequency $\Omega_s = \Omega/2$ ($= 1.843$). The double pendulum swings subharmonically with a period $6T$ ($T = 2\pi/\Omega$). Apparently the motion is similar for $A = 1.79$ [Fig.~\ref{PhasePlot1} (ii)] with increased angular displacements with $b \approx -\sqrt{2}$. However, the enlarged view of the limit cycle [inset of Fig.~\ref{PhasePlot1} (ii)] and the power spectrum [Fig.~\ref{PS1} (ii)] confirm it as a subharmonic oscillation of period $2T$. The angular displacement increases further for $A = 2.24$. However, the limit cycle has now two closely spaced orbits [inset in Fig.~\ref{PhasePlot1} (iii)]. The corresponding power spectrum [Fig.~\ref{PS1} (iii)] has now several equispaced peaks at an interval of $\Omega/4$. The largest peak is still at $\Omega/2$. This is due to a period-doubling bifurcation of subharmonic oscillations. The period of oscillation, in this case, is $4T$.
	
The phase portraits, which are shown in the middle row [Fig.~\ref{PhasePlot1} (iv)-(vi)], are for $\Omega = 1.779$ (very near $\Omega_{R,2}$) and for different values of $A$. For $A = 0.71$, the driving parameters fall in the  instability zone $H_1^{(1)}$ and just outside $S_1^{(2)}$ [Fig.~\ref{marginal1} (ii)].  The inset shows three closely placed lines. The power spectrum of $\theta_1$ [Fig.~\ref{PS1} (iv)] has several peaks, with one at the lowest frequency $\Omega/3$. The larger peaks are at an integral multiple of $\Omega$. Here, two masses show out-of-phase period-three harmonic motion [Fig.~\ref{PhasePlot1} (iv)]. As $A$ is raised to a value $1.5$, the driving parameters now fall in the overlap region of the instability zones labeled as $H_1^{(1)}$ and $S_1^{(2)}$ [Fig.~\ref{marginal1} (b)]. The out-of-phase period-three harmonic oscillations flip to in-phase subharmonic oscillations. Now, the limit cycles have a finite area in the $\theta_1$-$\theta_2$ plane. Further increase in $A$ makes the limit cycle asymmetric. For $A = 1.75$ the limit cycle does not have the inversion symmetry. The asymmetry in the limit cycle appears when the power spectrum shows comparable  peaks both at odd and even multiples of $\Omega/2$. The oscillations, however, remain subharmonic. The limit cycles shown in the lower row [Figs.~\ref{PhasePlot1} (vii)-(ix)] are for driving frequency $\Omega  = 1.512$ (near $\Omega_{R,3}$). The driving parameters are inside the  instability zone $S_1^{(2)}$ and outside $H_1^{(1)}$. The oscillations are in-phase for all three values of $A$ considered here. For $A = 0.54$, the projection of the limit cycle in the configuration space is almost like a curve. The power spectrum of $\theta_1$ [Fig.~\ref{PS1} (vii)] for this case shows the motion to be a period-four subharmonic oscillation. As $A$ is increased, the limit cycles enclose finite area in the configuration space. The pendulum shows period-three and period-five oscillations for $A = 0.76$ and $1.04$ [Fig.\ref{PhasePlot1} (viii) and (ix)] respectively, as evident from the corresponding insets of these two plots, which is further supported by the corresponding power spectra shown in  Fig.~\ref{PS1} (viii)-(ix). 
	
Figures~\ref{PhasePlot2} (i)-(ix) show the phase portraits in the configuration space of a double pendulum with two unequal masses ($\mu = 1/2$). Limit cycles are shown for different values of $\Omega$, $A$ with  $\beta = 0.1$. The plots in the top, middle and bottom rows are for $\Omega = 3.057$, $2.013$ and $1.533$, respectively. These frequencies are very close to the three extrema of the merged marginal curve for $\beta = 0.1$ [see Fig.~\ref{marginal_curves} (iii)].

\begin{figure}[!]
	\centering
	\includegraphics[height=!, width=16cm] {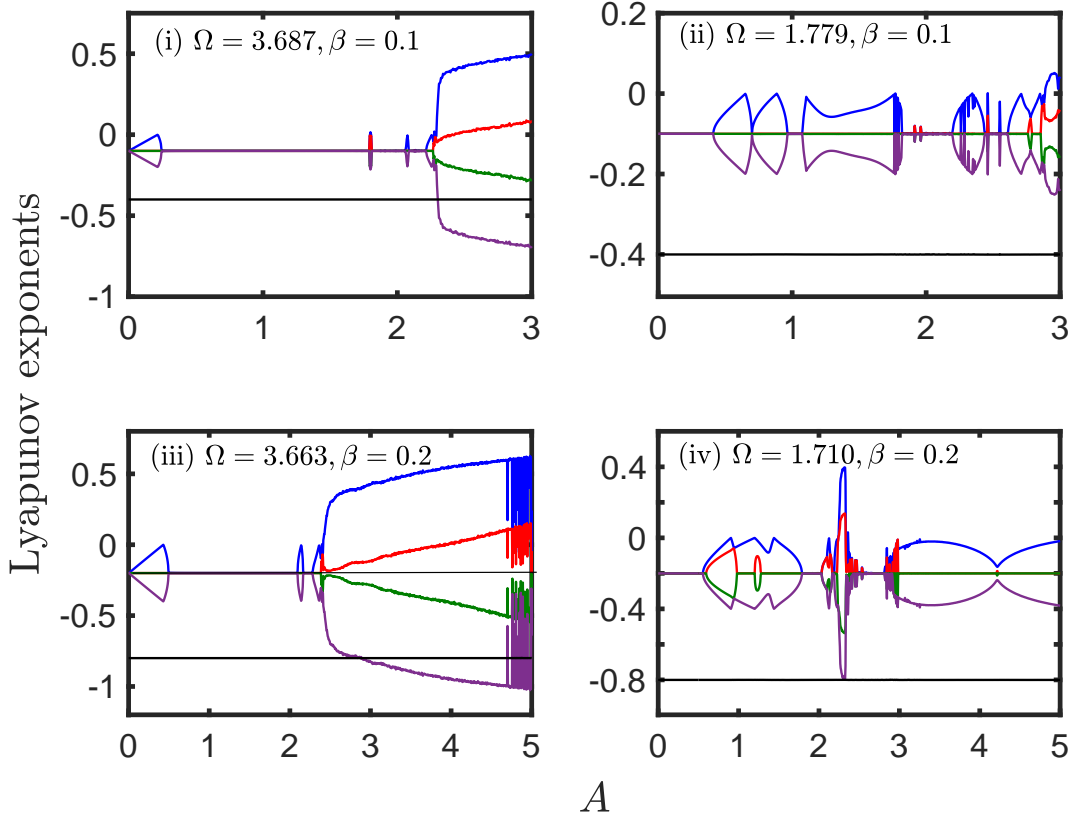}
	\caption{(Colour online) The variation of the four Lyapunov exponents [shown with coloured (gray) lines] as a function of driving amplitude $A$ for different values of the driving frequency $\Omega$ and damping coefficient $\beta$: (i) $\Omega = 3.687$, 
		$\beta = 0.1$; (ii) $\Omega = 1.779$, $\beta = 0.1$; (iii) $\Omega = 3.663$, $\beta = 0.2$; (iv) $\Omega = 1.710$, $\beta = 0.2$. Other parameters are: $\mu = 1$ and $\lambda = 1$. The black coloured straight line in each view graph is the sum 
		of all the Lyapunov exponents.} \label{Lyapunov_exponent}
\end{figure} 

\begin{figure}[!]
	\centering
	\includegraphics[height=!, width=16cm] {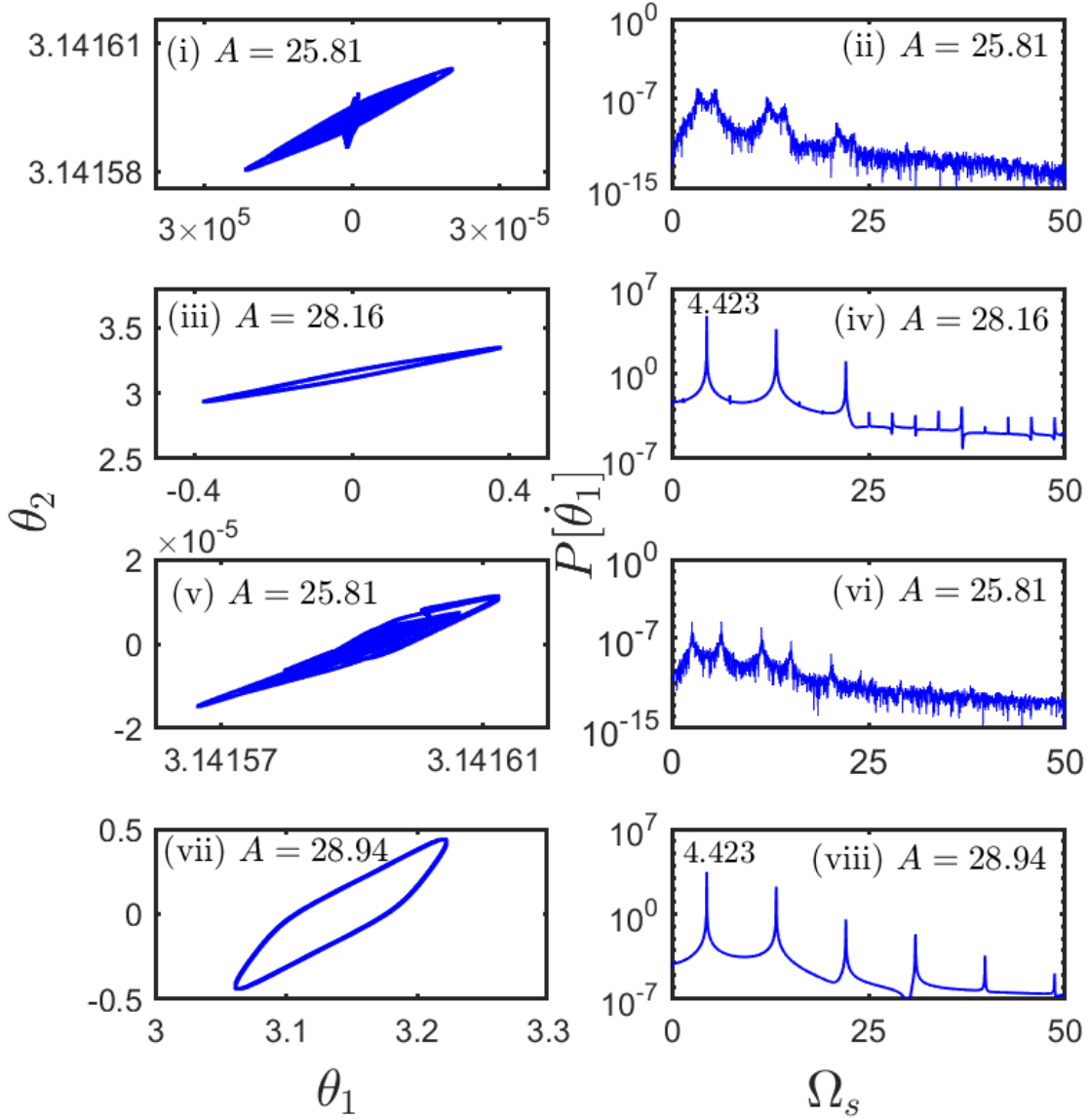}
	\caption{(Colour online) 
		Stabilization of a partially inverted driven coplanar double pendulum for $\mu = 0.5$, $\lambda = 1$, $\beta = 0.1$ and $\Omega = 8.845$.	Phase portraits in the $\theta_1$-$\theta_2$ plane (left column) and power spectra of $\dot{\theta}_1$ (right column)   are plotted for different values of  $A$.} \label{half_inverted}
\end{figure} 

\begin{figure}[!]
	\centering
	\includegraphics[height=!, width=16cm] {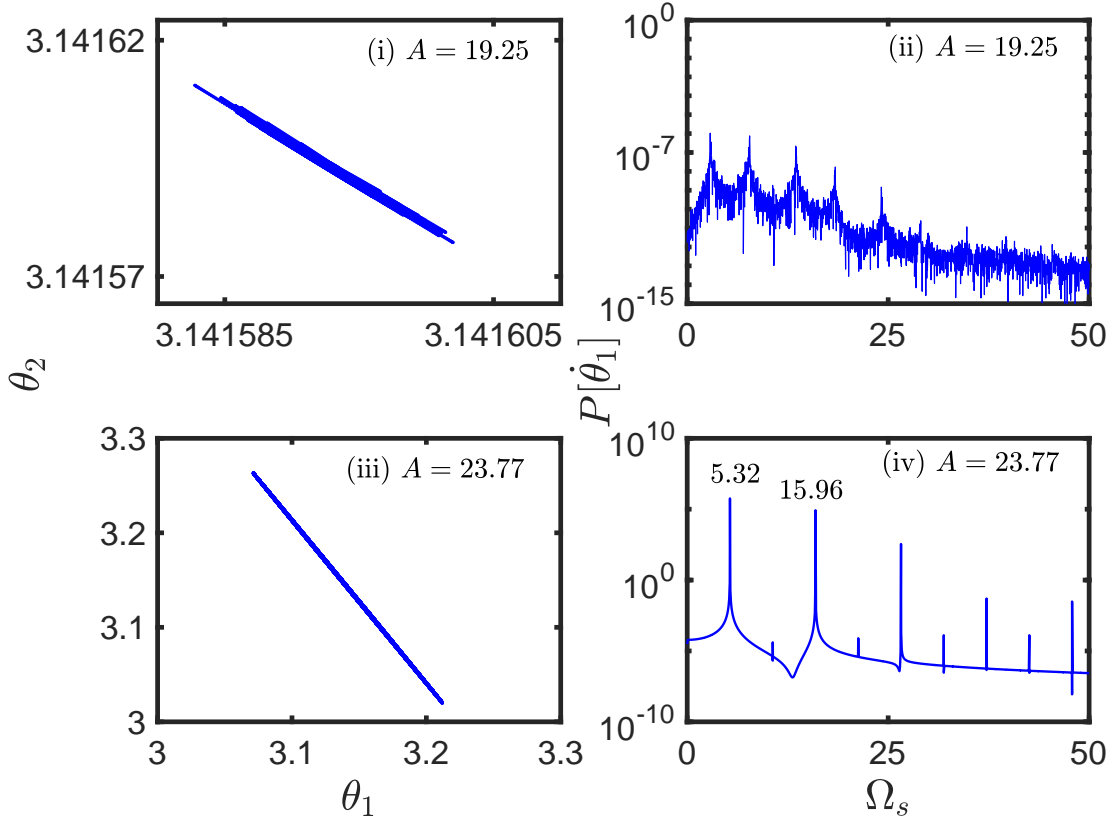}
	\caption{(Colour online) Stabilization of a fully inverted driven coplanar double pendulum for $\mu = 0.5$, $\lambda = 1$, $\beta = 0.1$ and $\Omega = 10.64$. Phase  portraits in the $\theta_1$-$\theta_2$ plane and the corresponding power spectra of the phase variable $\dot{\theta}_1$ for  $A = 19.25$ are  given the top row, while those for  $A = 23.77$ are in the bottom row.} \label{fully_inverted}
\end{figure} 

Inset in each plot shows an enlarged view of a small part of the corresponding limit cycle in the $\theta_1$-$\dot{\theta}_1$ plane. For all three values of $A$ ($0.34$, $0.90$, and $1.80$) at $\Omega = 3.057$ (the top row),  the pendulum shows out-of-phase oscillations. For $A=0.34$, which is just above the first minimum of the merged subharmonic instability zone, the limit cycle apparently looks as an ordinary subharmonic case. However, a closer look reveals a period-nine subharmonic oscillations [Inset of Fig.~\ref{PhasePlot2} (i)]. The period-nine oscillations are confirmed by the power spectrum of $\theta_1$ [Fig.~\ref{PS2} (i)], which has interesting features. The first tiny peak appears at $\tilde{\Omega} = \Omega/9$. Subsequent smaller peaks are at $(2\tilde{m}+1) \tilde{\Omega}/2$, where $\tilde{m}$ is a natural number. Larger peaks appear at $(2m+1)\Omega/2$. Inside the primary (dominating) subharmonic peak at $\Omega/2$, there are secondary subharmonic peaks at $(2\tilde{m}+1) \tilde{\Omega}$. As $A$ is raised further, the nature of oscillations changes. For $A = 0.90$ ($1.80$) [Fig.~\ref{PhasePlot2} (ii) (Fig.~\ref{PhasePlot2} (iii))], the pendulum shows ordinary (period-two) subharmonic oscillations. The first two limit cycles, which are plotted in the middle row [Figs.~\ref{PhasePlot2} (iv) and \ref{PhasePlot2}(v)], are for $A= 1.55$ and $1.58$, which are just above the cusp  of the merged marginal curve [Fig.~\ref{marginal_curves}(iii)]. The limit cycles are qualitatively different from those shown in the top row. The limit cycles show period-five and period-three subharmonic oscillations. The power spectra (not shown here) confirm these observations. The oscillation periods are very sensitive to the values of $A$ at this driving frequency. With further increase in $A$, the driving parameters fall inside the upper region of the merged subharmonic instability zone. The pendulum oscillates subharmonically [Fig.~\ref{PhasePlot2} (vi)]. The first limit cycle plotted in the bottom row [Fig.~\ref{PhasePlot2} (vii)] is for $\Omega = 1.533$ and $A=0.911$, which is just above the second minimum of the merged subharmonic marginal curve. The pendulum shows in-phase  period-five subharmonic oscillations [Inset of Fig.~\ref{PhasePlot2} (vii)], which are confirmed by the power spectrum of $\theta_1$ [Fig.~\ref{PS2} (ii)]. The first tiny peak appears at $\tilde{\Omega} = \Omega/5$ and the subsequent smaller peaks are at $(2\tilde{m}+1) \tilde{\Omega}/2$, as described earlier for Fig.~\ref{PS2} (i). Larger peaks appear at $(2m+1)\Omega/2$. The ordinary subharmonic oscillations are restored for slightly higher values of $A$ [Figs.~\ref{PhasePlot2} (viii) and \ref{PhasePlot2}(ix)].
\begin{figure*}[!]
	\centering
	\includegraphics[height=!, width=17cm] {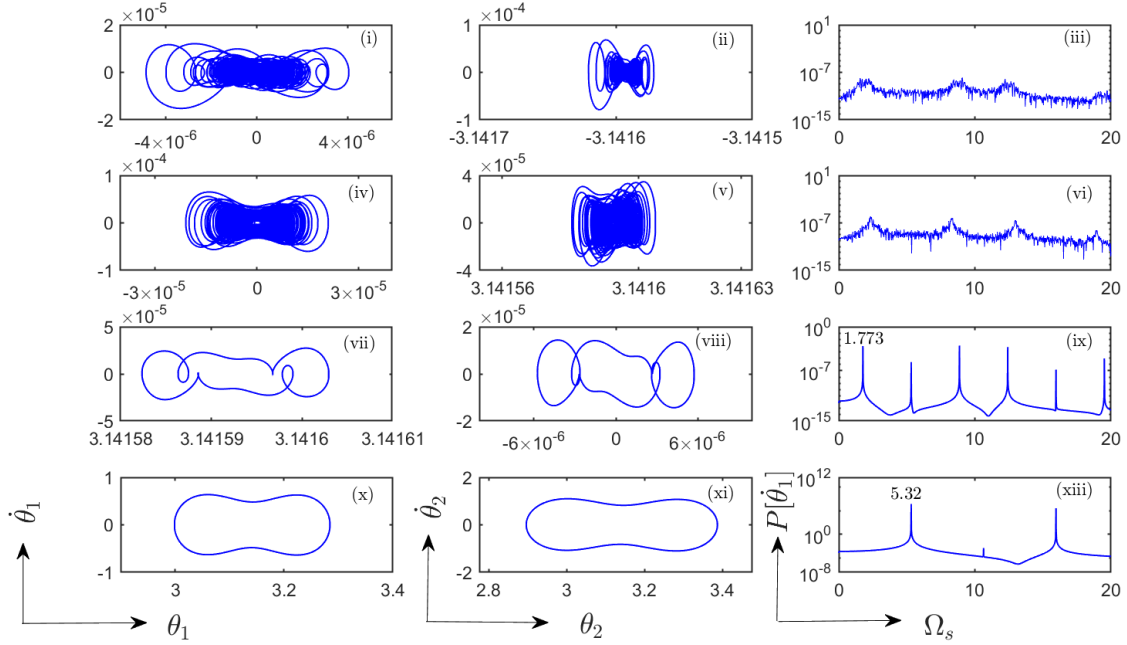}
	\caption{(Colour online) Phase portraits and power spectra for $\mu = 0.5$, $\beta = 0.1$, $A = 24.90$ and $\Omega = 10.64$. Phase portraits in the $\theta_1$-${\dot\theta}_1$ plane (column 1), the $\theta_2$-$\dot{\theta}_2$ plane (column 2) and the corresponding power spectra for phase variable $\dot\theta_1$ (column 3) are shown for different sets of initial conditions.} \label{inverted_diff_int}
\end{figure*} 

\begin{figure}[!]
	\centering
	\includegraphics[height=!, width=16cm]
	 {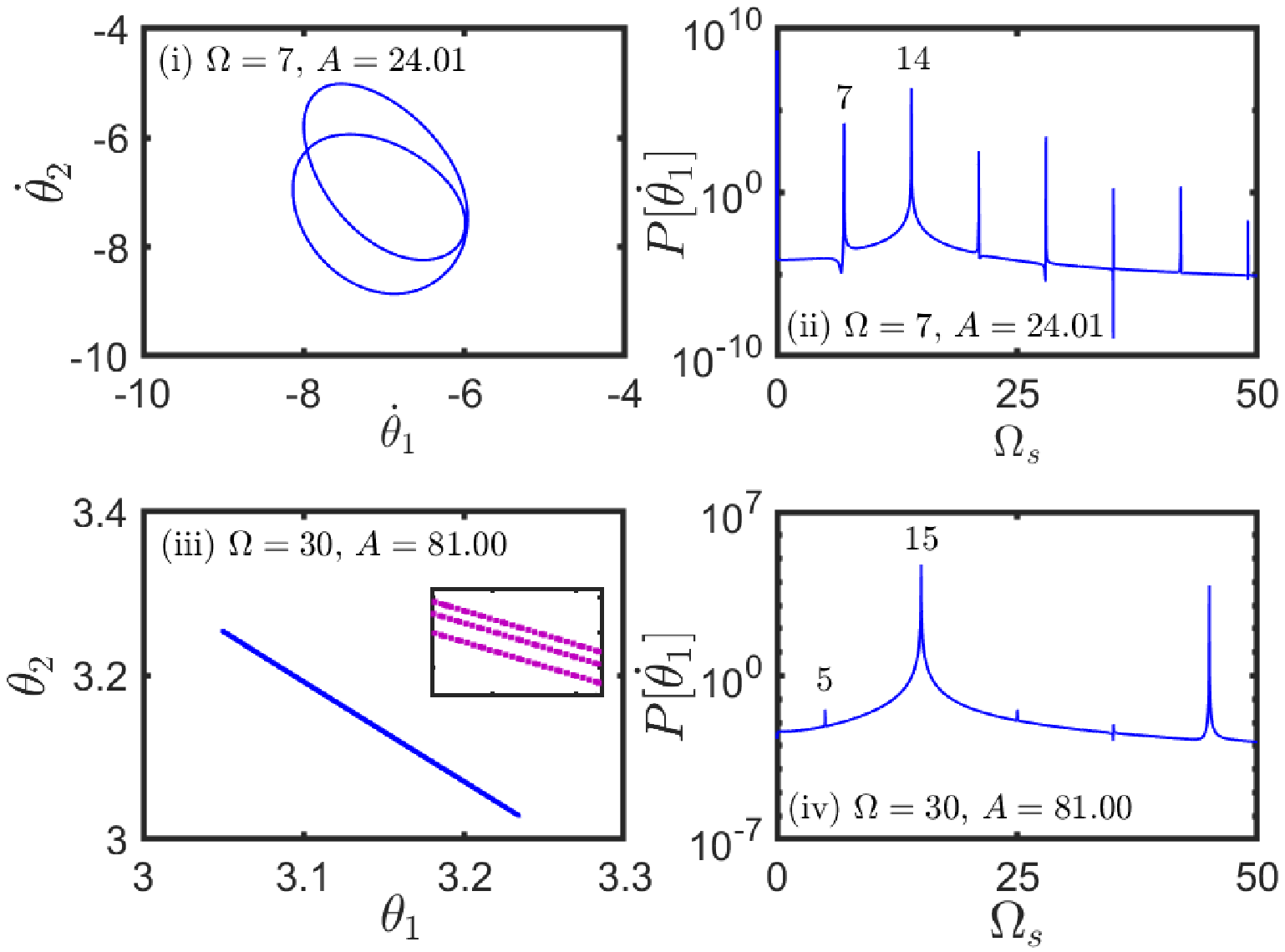}
	\caption{(Colour online)
	Rotational and oscillatory motion of a driven coplanar double pendulum for $\beta = 0.1$. The phase portrait in $\dot\theta_1$-$\dot\theta_2$ plane [(i)] and the power spectrum for the phase variable $\dot\theta_1$ [(ii)] are for a rotational motion with $\mu = 0.1$, $\Omega = 7$, $A = 22.54$. Similarly, the phase portrait in the $\theta_1$-$\theta_2$ plane [(iii)] and the power spectrum for the variable $\dot{\theta}_1$ [(iv)] are for an oscillatory motion for $\mu = 2$, $\Omega = 30$, $A = 81.00$. Inset in (iii) shows a zoomed view of a tiny portion of of limit cycle in the $\theta_1$-$\theta_2$ plane.} \label{rotation_swing}
\end{figure} 
	
Figure~\ref{PhasePlot3} displays the large amplitude oscillatory and rotational motion of a coplanar double pendulum. Limit cycles are shown in the velocity space [$\dot{\theta}_1$ - $\dot{\theta}_2$ ($X_2$-$X_4$) plane] at different driving frequencies for two different values of driving amplitude. The top row shows the limit cycles for $A = 2.20$ and at four different values of $\Omega$. Power spectra of the angular velocity $\dot{\theta}_1$ for four different frequencies are shown in the second row. For $\Omega = 3.057$, the limit cycle [Fig.~\ref{PhasePlot3} (i)] is a multiple orbit closed curve. The power spectrum of the angular velocity $\dot{\theta}_1$ for the limit cycle shown in Fig.~\ref{PhasePlot3} (i) is given in Fig.~\ref{PhasePlot3} (v). The peaks in the power spectrum occur at the integral multiples of $\Omega/12$ with the largest one at the frequency of $\Omega/2 = 1.528$. The oscillations are subharmonic of period-six. For $\Omega = 2.413$, the limit cycle [Fig.~\ref{PhasePlot3} (ii)] is quite different. The corresponding power spectrum $P[\dot{\theta}_1]$ [see, Fig.~\ref{PhasePlot3} (vi)] confirms a subharmonic response. At $\Omega = 2.013$, the pendulum motion is rotational for neither $\dot{\theta}_1$ nor $\dot{\theta}_2$ changes sign. It is a case of harmonic (synchronous) rotation. For $\Omega = 1.6$, the pendulum shows period-two harmonic rotational motion. The corresponding power spectrum $P[\dot{\theta}_1]$ [Fig.~\ref{PhasePlot3} (viii)] shows larger peaks located at the integral multiple of $\Omega$, while the smaller ones are found at odd multiples of $\Omega/2$. The third row shows the limit cycles for the pendulum motion in the velocity space at different values of $\Omega$ for $A = 1.60$. For $\Omega = 3.057$, $2.413$ and $2.013$ [Fig.~\ref{PhasePlot3} (ix), Fig.~\ref{PhasePlot3}(x) and Fig.~\ref{PhasePlot3} (xi)] the pendulum oscillates subharmonically with respect to the driving. For $\Omega = 1.6$, the pendulum rotates chaotically. The corresponding phase portraits in the velocity space and the power spectra are shown in Figs.~\ref{PhasePlot3} (xii) and~\ref{PhasePlot3} (xvi) respectively.
	
Further increase in driving amplitude leads to chaotic motion. The Lyapunov exponent, $\zeta$, is a measure of separation between two neighboring trajectories that evolve in the phase space. As our dynamical system [Eqs.~(\ref{dynsys1}-\ref{dynsys5})] in the autonomous form is five-dimensional, there are five Lyapunov exponents ($\zeta_i$ for $i = 1,2, ..., 5$) for a parametrically excited double pendulum. Due to periodic driving, one of the five exponents is always zero. We have numerically computed the other four Lyapunov exponents as a function of driving amplitude $A$ for fixed values of other parameters, following the method by Wolf \textit{et al.}~\cite{Wolf_etal_1985}. When the largest Lyapunov exponent becomes definitely positive, the trajectories of a double pendulum move chaotically in its phase space. Figure~\ref{Lyapunov_exponent} displays the variations of other four Lyapunov exponents ($\zeta_i$, $i=1, 2, 3, 4$) as a function of $A$ for fixed values of $\Omega$ and $\beta$. The dimensionless driving amplitude $A$ is raised in small steps of $0.005$ up to a maximum value of $3.0$ ($5.0$) for $\beta = 0.1$ ($0.2$). Plots in the upper (lower) row are for $\beta = 0.1$ ($0.2$). The variations of the Lyapunov exponents with $A$ at the largest resonance frequency for subharmonic (harmonic) excitation are shown in the left (right) column. The largest Lyapunov exponent $\zeta_1$ becomes positive at $A = 2.290$ ($2.415$) for $\beta = 0.1$ ($0.2$) and $\Omega = \Omega_{R, 1} = 3.687$ ($3.663$), as shown in Fig.~\ref{Lyapunov_exponent} (i) (Fig.~\ref{Lyapunov_exponent} (iii)). The pendulum shows chaotic rotations about the point of support. The second largest Lyapunov exponent $\zeta_2$ also becomes positive for $A = 2.435$ ($4.05$) for $\beta = 0.1$ ($0.2$) and the rotational motion of the pendulum becomes hyperchaotic~\cite{Harrison-Lai_1999}. Figure ~\ref{Lyapunov_exponent} (ii) (Fig.~\ref{Lyapunov_exponent} (iv)) displays the variations of the exponents with $A$ for $\beta = 0.1$ ($0.2$) and $\Omega = \Omega_{R, 2} = 1.779$ ($1.710$). The chaotic rotations begin at $A = 2.905$ for $\beta = 0.1$. For the case of $\beta = 0.2$, the chaotic rotations are observed in a small window of $A$ ($2.25 < A < 2.33$). The sum of all the Lyapunov exponents~\cite{Das-Kumar_2015} is $-4 \beta$ for $\mu =1$, as shown by the black straight line in all the viewgraphs of Fig.~\ref{Lyapunov_exponent}.

\subsection{Motions of partially or fully inverted double}

Stabilization of an inverted single pendulum was first studied by Kapitza~\cite{Kapitza_1951}.  Interesting experiments on the stability of an inverted double pendulum were performed by Acheson and Mullin~\cite{Acheson-Mullin_1993,Mullin_etal_2003}.
We now discuss the effects of parametric driving on unstable states corresponding to a partially or fully inverted double pendulum with $\mu = 0.5$ and $\beta = 0.1$. Driving the pendulum at a frequency much larger than the largest resonance frequency $\Omega_{R, 1}$ has interesting results. Figure \ref{half_inverted} shows the stabilization of a partially inverted pendulum with driving frequency $\Omega = 8.845$ and for different values of $A$. The phase portraits in the $\theta_1$-$\theta_2$ plane are displayed in the left column, while the corresponding power spectra of the phase variable $\dot{\theta}_1$ are shown in the right. The top row
shows the results for $A = 25.81$, for which the parameter values ($\Omega, A$) are outside the subharmonic instability zone  but inside that for the harmonic instability
[Fig.~\ref{Inverted_double_stability_mu0,5_combined} (ii)]. Figure~\ref{half_inverted} (i) shows that the parametric driving stabilizes the fixed point ($0, \pi$). The corresponding power spectrum [Fig.~\ref{half_inverted} (ii)] shows insignificant noise and confirms the same. As $A$ is raised slightly above to a value $28.16$, the double pendulum shows small amplitude oscillations about the fixed point ($0, \pi$). The corresponding limit cycle is displayed in Fig.~\ref{half_inverted} (iii).  The power
spectrum [Fig.~\ref{half_inverted} (iv)] shows that the oscillatory motion of this partially inverted
system is subharmonic. For  $A = 25.81$, the fixed point ($\pi, 0$) may also be stabilized, as shown in [Figs.~\ref{half_inverted} (v) and \ref{half_inverted} (vi)]. As the driving amplitude is varied from $25.81$ to $28.94$, the fixed point ($\pi, 0$) bifurcated to a limit cycle, as displayed in Fig.~\ref{half_inverted} (vii). The power spectrum of $\dot{\theta}_1$ shows again subharmonic oscillations about ($\pi, 0$).

Figure~\ref{fully_inverted} shows the results for a fully inverted double pendulum under parametric driving.  The double pendulum may be stabilized in fully inverted state for $\Omega = 10.64$ and $A = 19.25$ [see Fig.~\ref{fully_inverted} (i)], which is confirmed by its power spectrum  [Fig.~\ref{fully_inverted} (ii)]. Increasing $A$ to a value $23.77$, it shows stable small amplitude oscillations about the inverted position. The limit cycle corresponding to the small oscillations about the fixed point ($\pi, \pi$) is shown in Fig.~\ref{fully_inverted} (iii). The corresponding power spectrum, Fig.~\ref{fully_inverted} (iv), confirms that the oscillatory motion is subharmonic. It is observed that a double pendulum may be maintained in a steady state corresponding to a partially or a fully inverted position, if the driving frequency is considerably larger than $\Omega_{R, 1}$. 

For sufficiently large driving amplitude $A$ and frequency $\Omega$, the instability zones for different fixed points overlap, and the resultant stable and steady motion of the double pendulum is sensitive to the initial conditions. Fig.~\ref{inverted_diff_int} displays different possibilities for $\mu = 0.5$, $\beta = 0.1$, $\Omega = 10.64$ and $A = 24.90$ depending upon the initial conditions. The driving parameters ($\Omega$, $A$) in this case are simultaneously inside the subhmarmonic instability zones for the normal state ($0$, $0$) and fully inverted state ($\pi$, $\pi$) [see, Fig.~\ref{Inverted_double_stability_mu0,5_combined} (i) and Fig.~\ref{Inverted_double_stability_mu0,5_combined} (iv)]. However, they are inside the harmonic instability zones for partially inverted states [($0$, $\pi$) and ($\pi$, $0$)] as shown in Fig.~\ref{Inverted_double_stability_mu0,5_combined} (ii) and Fig.~\ref{Inverted_double_stability_mu0,5_combined} (iii).  The results shown in the four rows (top to bottom) are for four different initial conditions ($\theta_1$, $\dot{\theta}_1$,  $\theta_2$,  $\dot{\theta}_2$  at $\tau  = 0$).  The projections of the phase portraits in the $\theta_1$-$\dot{\theta}_1$ and in  the $\theta_2$-$\dot{\theta}_2$  planes are plotted for different initial conditions in the first and second columns, respectively. The power spectrum of $\dot{\theta}_1$ for the corresponding case is plotted in the third column. The results obtained with the initial conditions ($0.0175$, $0.0$, $0.0175$, $0.0$) are shown in the top row [Fig.~\ref{inverted_diff_int} (i)-(iii)]. Similar results are obtained [Fig.~\ref{inverted_diff_int} (iv)-(vi)] with initial conditions ($0.0175$, $-0.1$, $3.130$, $0.1$). They correspond to the stabilization of the fixed point ($0$, $\pi$). Using the initial conditions ($3.130$, $0.1$, $0.0175$, $-0.1$) leads to the stabilization of the other partially inverted state ($\pi$, $0$) [Figs.~\ref{inverted_diff_int} (vii)-(viii)]. The corresponding power spectrum [Fig.~\ref{inverted_diff_int} (ix)] shows the first peak near response frequency $\Omega/6$, although the heights of the peaks are insignificant.  For the same parameter values but a different set of initial conditions ($3.150$, $-0.1$, $3.131$,  $0.1$)  leads to a limit cycle about the fully inverted state ($\pi$, $\pi$), as shown in Fig.~\ref{inverted_diff_int} (x) and \ref{inverted_diff_int} (xi). The power spectrum for this case [Fig.~\ref{inverted_diff_int} (xii)] shows the limit cycle corresponds subharmonic motion of the double pendulum.

We now consider two cases: (1) $\mu = 0.1 (\ll 1)$, $\Omega = 7$, $A = 22.54$ and (2) $\mu = 2.0$, $\Omega = 30$, $A = 81.00$. The dimensionless damping coefficient $\beta = 0.1$  is the same for two cases. Figure~\ref{rotation_swing} (i) shows the phase portrait of a double pendulum in the $\dot{\theta}_1$-$\dot{\theta}_2$ plane for the first case.  The limit cycle corresponds to a rotational motion of the pendulum. The angular velocities vary periodically although the angular displacements keep increasing. The corresponding phase portraits in the $\theta_1$-$\theta_2$ plane are not shown here. The power spectrum has several peaks [Fig.~\ref{rotation_swing} (ii)]. The first peak is located at the dimensionless $\Omega_s = 7$, while the largest peak is located at response frequency twice of that.  The limit cycle corresponding to this harmonic rotational motion consists of two loops in the $\dot{\theta_1}$-$\dot{\theta_2}$ plane. Effectively, the limit cycle is synchronous with the driving (period $T$). Now we consider a case for which $\mu > 1$. Figure \ref{rotation_swing} (iii) displays the phase portrait for $\mu = 2$, $\beta = 0.1$,  $\Omega = 30$, and $A = 81.00$ in the $\theta_1$-$\theta_2$ plane. The motion is periodic motion about the fully inverted state ($\pi$, $\pi$). The corresponding limit cycle is squeezed to a straight line in the $\theta_1$-$\theta_2$ plane. This can happen when $\theta_2$ is proportional to $\theta_1$. A closer look suggests the limit cycle corresponds to subharmonic oscillations of period-3. A zoomed view of a portion of the limit cycle has three lines [see the inset, Fig.~\ref{rotation_swing} (iii)].  The power spectrum of the phase variable $\dot{\theta}_1$ shows the largest peak at response frequency equal to $\Omega/2$. The first peak, which is very small, is located at $\Omega/6$. We also note that a fully inverted double pendulum may be maintained in a steady state under parametric periodic driving at larger values of the driving frequency.

\bigskip	

\section{Conclusions}

We have investigated the motion of a damped coplanar double pendulum when the pivot oscillates sinusoidally with dimensionless amplitude $A$ and frequency $\Omega$. This problem is of general interest and applicable to a variety of physical systems in diverse areas of science and engineering.  The linear stability analysis shows the $\Omega$-$A$ plane is divided into several tongue-shaped subharmonic and harmonic instability zones for each normal mode of the pendulum. The marginal stability curves corresponding to any one of the normal modes, never intersect. However, the zones belonging to two different normal modes may overlap. Depending on the values $\Omega$ and $A$, a double pendulum may be excited to oscillate or rotate either subharmonically or harmonically.  If the driving parameters are chosen from a region of subharmonic and harmonic instability zone overlap with random initial conditions, then the pendulum oscillates or rotates subharmonically. A double pendulum has two pairs of Floquet multipliers. The Floquet multipliers intuitively help to understand the variety of motions executed by a double pendulum.  A new phenomenon is observed if the two masses of a double pendulum are unequal in the presence of moderate damping. Two zones for subharmonic instability belonging to two different normal modes merge together and form a new double-well-shaped instability zone with a barrier in between. This feature is unusual for a double pendulum and is observed for the first time to the best of our knowledge. The curvatures at the extrema of the new marginal curve are modified due to the merger. The linear stability analysis also predicts the possibility of stability of situations when only one of the pendulums is inverted. The Floquet analysis helps to unfold the mysteries of possible periodic solutions of a parametrically driven double pendulum.  

The nonlinear motion is investigated by constructing a dynamical system for the double pendulum. The limit cycles for a double pendulum with two equal masses corresponding to in-phase and out-of-phase oscillations are squeezed to a line in the configuration and velocity spaces. Such solutions, as predicted by the linearized equations of motion, continue in the nonlinear regime. A double pendulum may show multi-period oscillations near the instability onset. The power spectra of a phase variable reveal interesting features in these cases. For driving parameters in the pure harmonic (subharmonic) zones, the power spectra of a phase variable have larger peaks at frequencies equal to even (odd) multiples of $\Omega/ 2$. In the case of pure harmonic oscillations or rotations, smaller peaks are absent at frequencies which are less than $\Omega$. For period-$\nu$ ($\nu = 2, 3, 4, \cdots$) harmonic oscillations, smaller peaks are found at equal intervals of $\tilde{\Omega} = \Omega/\nu$ with the largest peak at $\Omega$. For the power spectra of period-$\nu$ subharmonic oscillations or rotations, smaller peaks are found at either odd or even multiples of $\tilde{\Omega}/2$. The corresponding largest peaks for both these period-$\nu$ solutions are found at $\Omega/2$. The periodic rotations of a double pendulum show rich dynamics, including period-doubling bifurcation and chaotic motion. The rotations become hyperchaotic for higher values of driving amplitude when the two largest Lyapunov exponents become positive. It is shown that by driving with relatively large amplitude and frequency for appropriately chosen initial conditions, the system could be made stable in partially or fully inverted states. The double pendulum may also oscillate in a partially or fully inverted state. This analysis will be useful for several problems where governing equations are primarily coupled-Mathieu equations with or without nonlinearity  and damping.

	

\end{document}